\documentclass[pre,aps,reprint,superscriptaddress]{revtex4-1}

\usepackage{graphicx}
\usepackage{amsfonts,amssymb}
\usepackage{epstopdf,epsfig}
\usepackage{bm}
\usepackage{bbold}
\usepackage{color}
\usepackage[nice]{nicefrac}
\usepackage{mathtools}
\usepackage{algorithm}
\usepackage{algpseudocode}
\usepackage{hyperref}

\graphicspath{{Figs/}}

\newcommand{\bX}{{\ensuremath\boldsymbol{X}}}

\newcommand{\bu}{\bm u}
\newcommand{\bF}{\bm f}

\newcommand{\pit}{\hat{\pi}_{\boldsymbol{\theta}}}
\newcommand{\circl}[1]{{\textcircled{\raisebox{-.9pt}{#1}}}}

\newcommand{\inria}{Universit\'e C\^ote d'Azur, Inria, CNRS, Sophia-Antipolis, France}
\newcommand{\cemef}{Ecole Nationale Sup\'erieure des Mines de Paris, PSL University, CNRS, Cemef, Sophia-Antipolis, France}

\begin{document}

\title{Steering undulatory micro-swimmers in a fluid flow through reinforcement learning}

\author{Zakarya El Khiyati}
\affiliation{\inria}
\author{Rapha\"{e}l Chesneaux}
\affiliation{\cemef}
\author{La\"etitia Giraldi}
\affiliation{\inria}
\author{J\'er\'emie Bec}
\affiliation{\inria}
\affiliation{\cemef}

\begin{abstract}
This work aims at finding optimal navigation policies for thin, deformable micro\-swimmers that progress in a viscous fluid by  propagating a sinusoidal undulation along their slender body. These active filaments are embedded in a prescribed, non-homogeneous  flow, in which their swimming undulations have to compete with the drifts, strains, and deformations inflicted by the outer velocity field. Such an intricate situation, where swimming and navigation are tightly bonded, is addressed using various methods of reinforcement learning.  Each swimmer has only access to restricted information on its configuration and has to select accordingly an action among a limited set. The optimisation problem then consists in finding the policy leading to the most efficient displacement in a given direction. It is found that usual methods do not converge and this pitfall is interpreted as a combined consequence of the non-Markovianity of the decision process, together with the highly chaotic nature of the dynamics, which is responsible for high variability in learning efficiencies. Still, we provide an alternative method to construct efficient policies, which is based on running several independent realisations of $Q$-learning.  This allows the construction of a set of admissible policies whose properties can be studied in detail and compared to assess their efficiency and robustness.
\end{abstract}

\maketitle

\section{Introduction}
\label{sec1}
A number of microorganisms, including bacteria and plankton, are natural examples of active, self-propelled particles. They often inspire the design of artificial devices used for industrial micro-manufacturing, toxic waste disposal, targeted drug delivery and localised medical diagnostics~\cite{wu2020medical}. Recent technological developments in the use of micro-swimmers in medicine open new frontiers, such as microscopic-scale surgery directly inside the human body and medicine and drugs delivery in very precise places where their efficiency will be optimal. Much work has been devoted to designing adequate nano-robots and studying the way they can be propelled and controlled using an external magnetic field~\cite{servant2015controlled}, in particular for \textit{in-vivo} conditions. Still, many questions remain open on how to optimise the displacement of these micro-swimmers, and in particular whether their behaviour is altered when they are embedded in complex flows comprising obstacles, walls, or having non-Newtonian properties. This is particularly important to design new strategies that will allow artificial swimmers to reach today inaccessible regions of the human body.

Studying and optimising the movement of swimmers and micro-swimmers is generally addressed in two successive steps. The first is to find an appropriate \textit{swimming strategy} by choosing the composition, shape, or deformation that will lead to an efficient locomotion. The second step is to define a \textit{navigation strategy} that takes into account obstacles, fluctuations in the surrounding flow, and its geometry, with the aim to minimise the time needed or the energy used to reach a specific target. Studying swimming strategies at the microscopic level requires advanced tools to describe fluid-structure interactions~\cite{berti2020swimming,alouges2013self}, to take a non-Newtonian rheology of the surrounding fluid into account~\cite{shen2011undulatory}, to model the hydrodynamics stresses due to the vicinity of walls~\cite{daddi2021hydrodynamics}. Finding an effective strategy then relies on biomimetics~\cite{borazjani2009numerical,cohen2010swimming} or on solving costly problems of optimal control~\cite{alouges2019energy}. As a matter of fact, such swimming issues are most of the time addressed in situations where the surrounding flow is at rest. This is justified by the complexity and the computational costs that would be required to accurately model the intricate fluid-structure interactions occurring in a chaotic or turbulent medium. 

Regarding navigation problems, there is an increasing interest in considering complicated carrier flows (see~\cite{vergassola2022ofactory} for a recent review). The swimming mechanisms are often oversimplified and one rather focuses on how to adjust macroscopic active features of the swimmers in order to optimise their long-term displacement. Under such conditions, the use of machine learning techniques has proved efficiency~\cite{cichos2020machine}. Reinforcement learning has for instance been used to address navigation in a turbulent flow and to construct strategies that allow swimmers to find optimal paths to their targets in such a chaotic and fluctuating environment~\cite{reddy2016learning,colabrese2017flow,gustavsson2017finding,schneider2019optimal,muinos2021reinforcement,gustavsson2022navigation,alageshan2020machine}. Navigation problems have also been studied from different perspectives such as finding new paths in the presence of obstacles that can be modelled as barriers of potential~\cite{schneider2019optimal}. As to approaches that use deep reinforcement learning, they demonstrated successes in various applications, such as terrain-adaptive dynamic locomotion~\cite{peng2016terrain} or real-world manipulation tasks~\cite{levine2016end}.

Here we want to address the situation where locomotion and navigation are tightly dependent on each other. Our goal is to show the feasibility of using machine learning approaches for a mesoscopic model of swimmer, and in particular to understand if such approaches are able, not only to make the swimmer move, but also to have it at the same time navigate a complex environment. The swimmers are assumed to be simple, deformable, inextensible thin filaments whose interactions with the fluid are explicitly described by the slender-body theory. Among the different types of swimming, we have chosen wave locomotion which is a self-propulsion strategy that relies on the generation and propagation of an undulation along the swimmer~\cite{pironneau1974optimal}. This is a relatively simple, but remarkably robust technique that builds on the interactions between the swimmer and the fluid and appears in a variety of swimming strategies observed in nature. We consider the problem where such swimmers are aiming at moving as fast as possible in a given direction, being at the same time embedded in a space-periodic, time-stationary, incompressible fluid flow that produces headwinds and deformations hindering their mobility.

We find that in such settings, the undulatory swimmers progress efficiently only if they follow a policy that prescribes different actions to be performed depending on their configuration. We focus on a simple, paradigmatic case: The actions and observations of the environment by the swimmer are both chosen from discrete sets that consist, respectively, of swimming either horizontally or vertically with different amplitudes and having sparse information on its orientation and the local direction of the flow. We look for optimal policies for this partially-observable Markov decision process, by applying and comparing various algorithms of reinforcement learning, ranging from the popular $Q$-learning technique to approximation methods (differential SARSA and Actor-Critic). We find that these approaches do not provide satisfactory results: Either they do not converge, or if they do so, they require prohibiting long times. We propose an alternative method that can be seen as belonging to the class of \textit{competitive} $Q$-learning approaches. It builds on the observation that, because of the highly chaotic character of the dynamics, individual realisations of simple, deterministic $Q$-learning are able to identify, quite quickly, a diversity of policies that lead to a reasonable displacement of the swimmer. The analysis of these admissible strategies can then be easily refined and systematised in order to rank them and select the most efficient ones. The advantage of this method is that it provides a short list of policies whose robustness can be tested and compared by varying the problem setting, for instance, the physical attributes of the swimmer (length, elasticity) or the properties of the surrounding flow.

The paper is organised as follows. Section~\ref{sec:model} introduces the swimmer model and reports results on how the efficiency of its locomotion depends on its physical properties. In Section~\ref{sec:statement}, we describe the outer flow and formulate the navigation problem in terms of discrete observations and actions. We also show that a policy is needed for the swimmer's displacement and introduce a naive strategy that allows it. Section~\ref{sec:reinforcement} is dedicated to a detailed comparison of various reinforcement learning techniques, leading to introduce the competitive $Q$-learning approach described above. Results on the performances and robustness of the short-listed policies are reported in Section~\ref{sec:robustness}, including trials performed in unsteady flows that are solving the Navier--Stokes equation. Finally, Section~\ref{sec:conclusion} gathers concluding remarks and perspectives.

\section{A model of undulatory threadlike swimmer}
\label{sec:model}
\subsection{Dynamics of deformable slender bodies} 
We consider elongated, flexible, inextensible swimmers. We moreover assume that they are very thin, meaning that their cross-section diameter $d$ is much smaller than their length $\ell$. This leads to describe their interactions with the surrounding viscous fluid in terms of the slender-body theory~\cite{lindner2015elastic}. The swimmers are embedded in an incompressible flow whose velocity field is denoted by $\bu(\boldsymbol{x},t)$. We neglect the swimmers feedback onto this prescribed flow, which is justified in the limit when swimmers are very thin and dilute. The conformation of an individual swimmer at time $t$ is given by a curve $\bX(s,t)$ parametrised by its arc-length $s\in[0,\ell]$. We neglect the swimmer's inertia, so that its dynamics is given by equating to 0 the sum of the forces that act on it, namely
\begin{align}
  - \zeta\,\mathbb{R}\left[\partial_t \bX-\bu(\bX,t)\right] &+\ \partial_s(T\partial_s \bX) \nonumber \\
  &- K\,\partial_s^4 \bX + \bF(s,t) = 0.
  \label{eq:vel_fib}
\end{align}
This equation of motion, which corresponds to the over-damped Cosserat equation, is the same as that obtained by resistive force theory to describe bio-filaments~\cite{moreau2018asymptotic}.  The first term on the left-hand side involves the drag coefficient $\zeta = 8\pi\mu/[2\log(\ell/d)-1]$ (with $\mu$ the fluid dynamic viscosity) and the local Oseen resistance tensor $\mathbb{R} = \mathbb{1} -(1/2)\,\partial_s\bX\,\partial_s\bX^{\mathsf{T}}$. This expression of the force exerted by the fluid assumes that, despite an arbitrary length, the fibre's thickness is so small that its perturbation on the flow has a vanishingly small Reynolds number, whence a linear but anisotropic drag. The second force appearing in Eq.~(\ref{eq:vel_fib}) is the tension. Its amplitude $T$  results from the inextensibility constraint $\vert\partial_s \bX (s,t)\vert = 1$, valid at all time $t$ and all position $s$ along the swimmer. The third term is the bending elasticity force and depends on the swimmer's flexural rigidity $K$ (product of Young's modulus and inertia). The last term, denoted by $\bF$, is a prescribed internal force that accounts for the \textit{active behaviour} of the swimmer responsible for its locomotion. Equation~(\ref{eq:vel_fib}) is associated with the free-end boundary conditions $\partial_s^2\bX(s,t) = 0$ and $\partial_s^3\bX(s,t) = 0$ at the swimmer's extremities $s=0$ and $\ell$. The tension itself satisfies a second-order differential equation obtained by imposing $\partial_t \vert\partial_s\bX\vert^2=0$ with the boundary conditions $T(s,t) = 0$ at $s=0$ and $\ell$.

In the absence of active force ($\bF=0$), the swimmer is just a passive, flexible but inextensible fibre, whose dynamics depends on two non-dimensional parameters. One is given by the ratio $\ell/L$ between the fibre's length $\ell$ and the characteristic spatial scale $L$ of the fluid flow.  It characterises to which extent the fibre samples the fluid flow length scales and monitors geometrical interactions with surrounding structures and eddies~\cite{picardo2018preferential,rosti2018flexible}. The other parameter is $(U\zeta/KL)^{1/4}\ell$, where $U$ is a typical magnitude of the fluid velocity. It measures the fibre's flexibility and in particular its likeliness to be bent or buckled by the flow~\cite{young2007stretch,brouzet2014flexible,allende2018stretching}. The larger it is, the more deformable is the fibre when it is subject to shear or compression.

\subsection{The undulatory swimming procedure}
\label{subsec:undulatory}
We focus on swimmers that move by propagating a sinusoidal plane wave along their body.  This undulation is assumed to be applied through the active body force $\bF$ appearing in the dynamical equation~(\ref{eq:vel_fib}). The swimmers are thus assumed to have the ability to adapt their curvature along their body, as in the case of nematodes~\cite{gray1964locomotion,berri2009forward}. Such settings are somewhat different from the beating of cilia or flagella, for which it is rather a time-periodic boundary data that is imposed to a flexible beating appendage, as in the case of sperm cells~\cite{friedrich2010high,jikeli2015sperm}. We choose here to write the active force as
\begin{equation}
    \bF(s,t) = A\,\zeta\,\nu\,\ell\,\cos(2\pi\,k\,s/\ell-\nu\,t)\,\bm p
    \label{eq:sinusoidal_force}
\end{equation}
where $\bm p$ is a unit vector in a direction orthogonal to that in which the swimmer is expected to move. The wave has frequency $\nu$ and wavenumber $2\pi k/\ell$ where $k$ is an integer.  To ensure self-propulsion, we impose that the force $\bF$ is not a global source of momentum for the swimmer, namely that $\int \bF \mathrm{d}s = 0$, justifying why the wavenumber has to be chosen as a multiple of $(2\pi/\ell)$. The strength of the active force is controlled by the dimensionless amplitude $A$. 

The resulting swimming speed in the $\bm p^\perp$ direction, which is hereafter denoted by $V_{\rm swim}$, non-trivially depends on the forcing parameters and the physical properties of the swimmer. To our knowledge, there is at present no analytic expression for $V_{\rm swim}$, even in the absence of external fluid flow ($\bu=0$).  This can be explained by the intricate role played by inextensibility and tension and the imposed free-end boundary conditions that prevent from obtaining an explicit solution for the fibre conformation $\bX$ for this force. Still, when rescaling spatial scales by the swimmer's length $\ell$ and time scales by the wave frequency $\nu^{-1}$, one finds that $V_{\rm swim} = \ell\nu\, \Psi_{k}(A,\mathcal{F})$, where $\mathcal{F} = (\zeta\nu/K)^{1/4}\ell$ is a non-dimensional measure of the swimmer's flexibility under the action of the active force and the $\Psi_k$'s are non-dimensional functions indexed by the wavenumber $k$. To obtain their behaviour, we have recourse to numerics.

To set our physical parameters and understand better how the swimmers respond to activation, we have  performed numerical simulations of the over-damped Cosserat equation~(\ref{eq:vel_fib}) for isolated fibres in a fluid flow at rest. We use the second-order, centred finite-difference scheme of~\cite{tornberg2004simulating} with $N=201$ to $801$ grid-points along the fibre's arc-length. The inextensibility constraint is enforced by a penalisation method. Time marching uses a second-order semi-implicit Adams--Bashforth method with time step ranging from $\delta t = 10^{-3}$ to $10^{-4}$. We have performed several simulations varying the forcing amplitude, its wavenumber, and the swimmer bending elasticity. After transients,  the swimmer, which is initialised in a straight configuration, develops an undulating motion corresponding to  a travelling wave propagating from its head ($s=0$) to its tail ($s=\ell$). Once this periodic regime is attained, we measure the time required for its displacement over several lengths $\ell$ in order to evaluate the asymptotic swimming speed $V_{\rm swim}$.

\begin{figure}[h]
\centering
\includegraphics[width=.7\columnwidth]{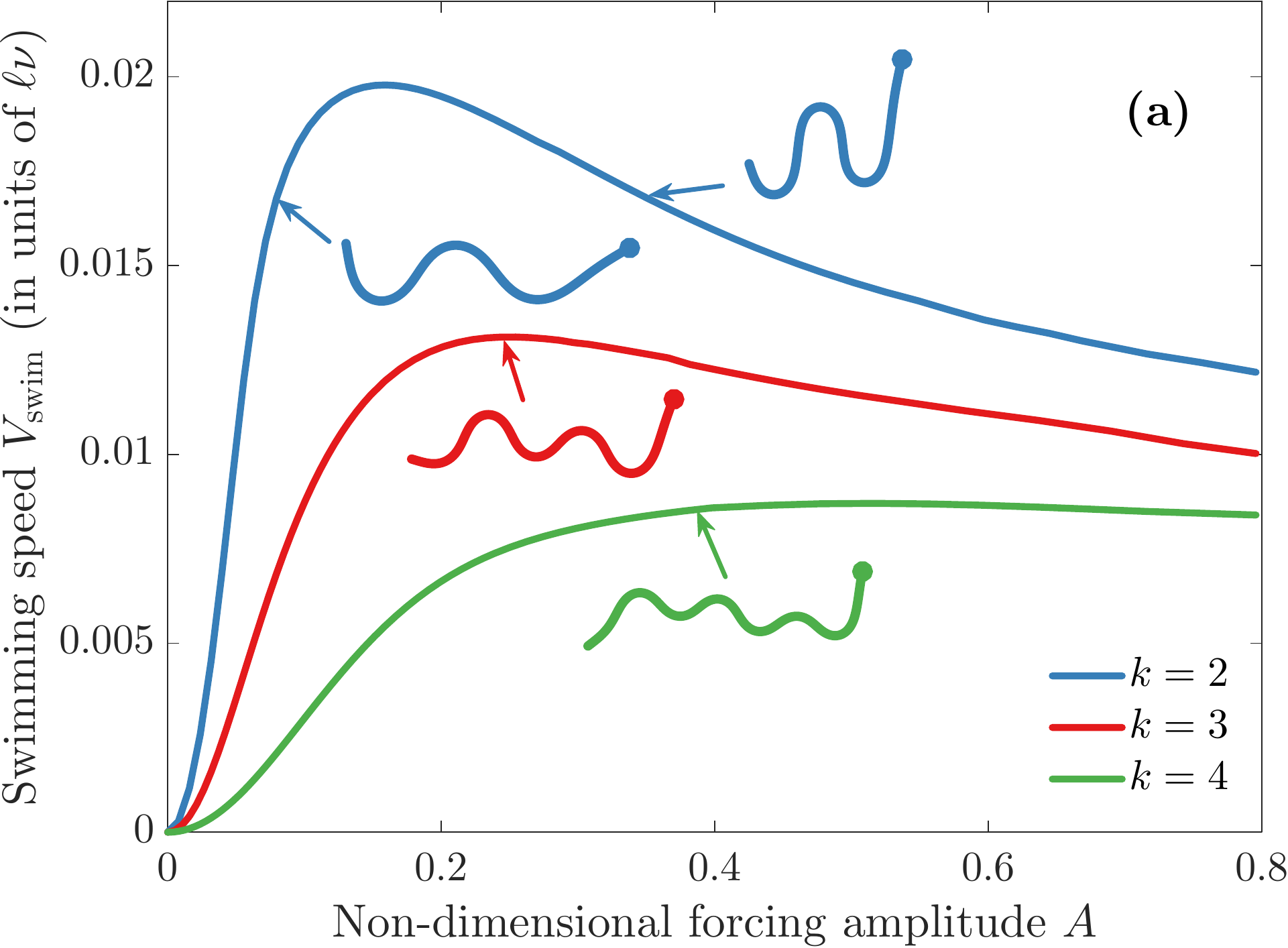}\\[8pt]
\includegraphics[width=.7\columnwidth]{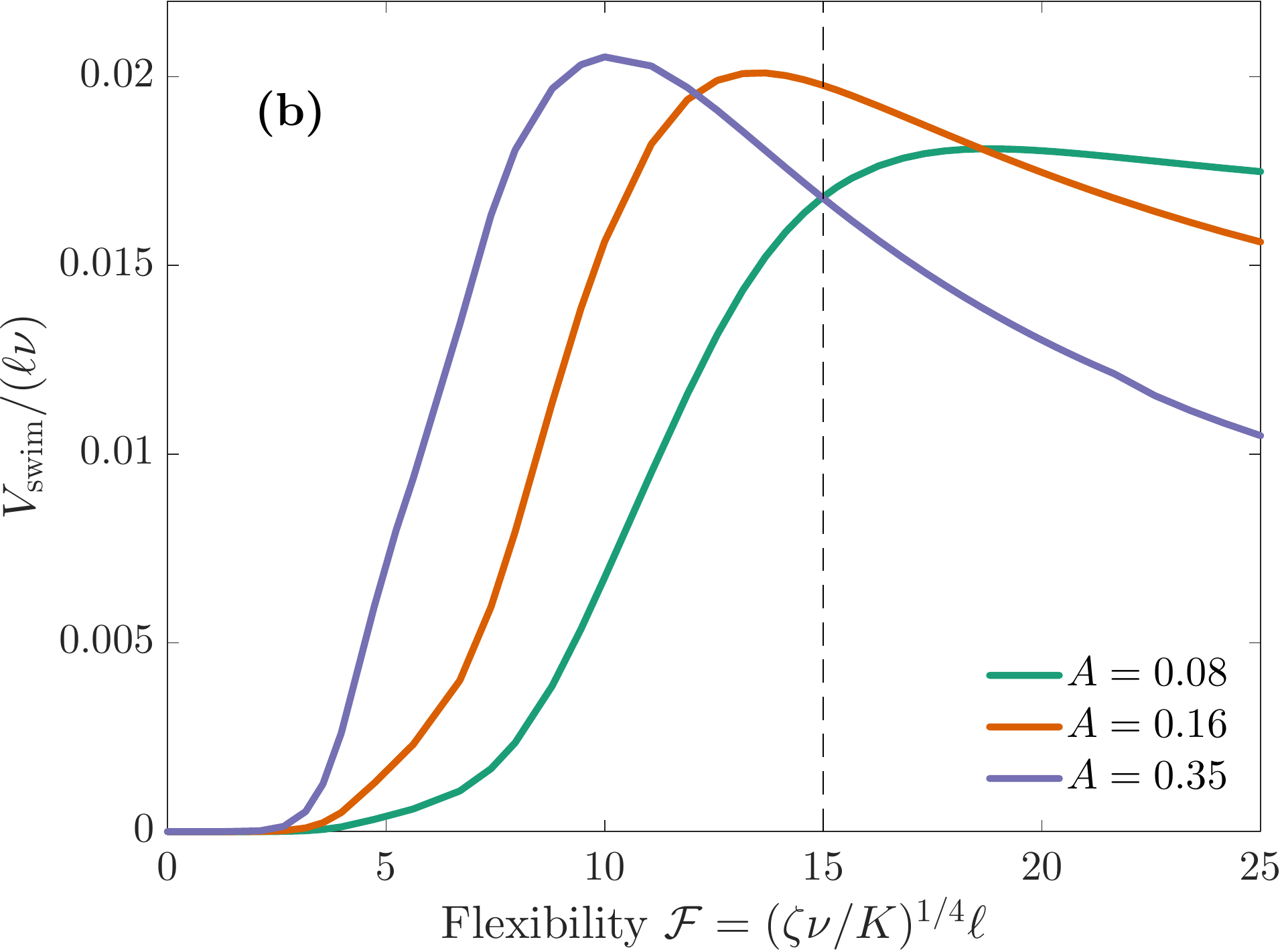}
\caption{\label{fig_vswim} Swimming speed in the absence of fluid velocity field, \textbf{(a)} as a function of the forcing amplitude, for flexibility $\mathcal{F}=15$ and three various values of the wavenumber ($k=2$, $3$, and $4$, as labelled), and \textbf{(b)} as a function of the swimmer's flexibility $\mathcal{F}$, for $k=2$ and three values of the forcing dimensionless amplitude $A$.}
\end{figure}

The dependence of the swimming speed upon the amplitude parameter $A$ is shown in Fig.~\ref{fig_vswim}(a), for different wave-numbers $k$ and a fixed dimensionless flexibility $\mathcal{F}$. Several representative configurations of the swimmer are also shown, with a dot indicating its head ($s=0$). At small forcing amplitudes, the undulation of the swimmer is very close to the imposed wave and the swimming speed increases quadratically. This behaviour can be obtained from a linear expansion of Eq.~(\ref{eq:vel_fib}) at $A\ll1$. To leading order the swimmer is aligned with $\bm p^\perp$, the unit vector orthogonal to the force, and it moves along this direction. The projection of its position can thus be expanded as $\bm p^\perp\cdot\bX = -s+X_1'$ with $X_1'\ll1$. In the transverse direction, one gets from Eq.~(\ref{eq:vel_fib}) that $X_2' = \bm p\cdot\bX = \mathcal{O}(A)$. The inextensibility constraint reads $\vert\partial_s\bX\vert^2 = (1-\partial_sX_1')^2 + (\partial_sX_2')^2 = 1$, implying that  the longitudinal perturbation $X_1'$ is of the order of $(X_2')^2$.  This indeed implies that $V_{\rm swim} \sim \partial_tX'_1 = \mathcal{O}(A^2)$. This quadratic growth saturates for $A\approx 0.1$--\,$0.2$ and the swimming speed then attains a maximum.  This optimal speed slowly decreases and shifts toward larger values of $A$ when $k$ increases. One consequently observes that achieving a given swimming speed is getting more energetic, or even impossible, when the wavenumber of the undulation is taken larger. Beyond this maximum, swimming becomes less and less efficient at larger forcing amplitudes. At such value the swimmer's distorsion is highly non-linear and bending elasticity becomes important and induces an important dissipation.

Figure~\ref{fig_vswim}(b) represents again $V_{\rm swim}$, but this time as a function of the non-dimensional flexibility $\mathcal{F}$, for $k=2$ and  three different amplitudes of forcing, before, at the maximum, and after.  The swimming speed attains a maximum at intermediate values of $\mathcal{F}$. When too stiff, the swimmer is not able to develop any significant undulation as the power input from the active force is dissipated by bending elasticity. At very large values of the flexibility, the swimmer is conversely too limp and energy is dissipated through viscous drag. An optimal locomotion is attained when the two dissipative mechanisms balance.

This preliminary study of undulatory swimming in the absence of an external flow allows us to properly choose the physical parameters that will be considered. Hereafter we focus on the forcing wavenumber $k=2$, the flexibility is chosen to be $\mathcal{F}=15$, and the forcing amplitudes are picked before the saturation of swimming efficiency, \textit{i.e.} $A\lesssim 0.15$.

\section{Statement of the navigation problem}
\label{sec:statement}
We consider the two-dimensional navigation problem, which consists in having the swimmer moving as fast as possible 
in the $x_1>0$ direction in the presence of a prescribed external fluid flow. In Sec.~\ref{subsec:cellflow}, after introducing the model flow, we demonstrate that displacement can only occur if the swimmer follows a strategy.  We then present in Sec.~\ref{subsec:optim} the observations and actions that can be used by the swimmers to control its displacement and we formulate the optimisation problem. We finally introduce in Sec.~\ref{subsec:naive} a ``naive'' strategy and evaluate its performance, with the aspiration that the reinforcement-learning methods applied in Sec.~\ref{sec:reinforcement} can outperform it.

\subsection{Swimming in a cellular flow}
\label{subsec:cellflow}

To design a navigation strategy, we consider an undulatory swimmer that is embedded is a two-dimensional cellular flow. More specifically, we prescribe the outer fluid velocity to be $\bm u = \nabla^\perp\Psi = (-\partial_2\Psi,\partial_1\Psi)$ with the stream function taking the simple periodic form $\Psi(\bm x,t) = (L\,U/\pi)\,\cos(\pi\,x_1/L)\,\cos(\pi\,x_2/L)$. The spatial domain is hence covered by a tile of cells mimicking eddies. Their size $L$ is chosen of the same order of magnitude as the fiber length $\ell$. The velocity field has an amplitude $U$ to be compared to the swimming velocity $V_{\rm swim}$ introduced in previous section. Such a two-dimensional flow is a stationary solution of the incompressible Euler equations and is used, for instance, to model the convection cells present in steady Rayleigh--B\'enard convection. It is often employed to study the effects of fluid shear and rotation on transport and mixing. It moreover has the convenience of being easily reproducible by experiments~\cite{rothstein1999persistent}. As seen later, even if the motion of tracers in such a flow is not chaotic, the dynamics of swimmers can be so.

Our aim is to maximise the swimmer displacement toward the $x_1>0$ direction. When using the basic swimming, that is to say always binding the fibre to swim with the force (\ref{eq:sinusoidal_force}) constantly applied along the direction $\bm p = \bm e_2$, one does not observe any long-term, net displacement.  We have indeed perform a set of numerical simulations where the swimmer is initialised in a straight configuration, with its head always oriented toward $x_1>0$, and varying its initial angle with the horizontal direction. Unless otherwise stated, we always use a discretisation of the swimmer with $N=201$ grid-points and a time step $\delta t =10^{-3}$.  Performance is then monitored by
\begin{equation}
	\bar{x}_1(t) = \bm e_1\cdot \bar{\bm X}(t) = \frac{1}{\ell}\int_0^\ell \bm e_1\cdot\bX(s,t)\,\mathrm{d}s,
\end{equation}
\textit{i.e.}\/ by the horizontal displacement of the swimmer's centre of mass $\bar{\bm X}$.

\begin{figure}[h]
  \centering
  \includegraphics[width=.7\columnwidth]{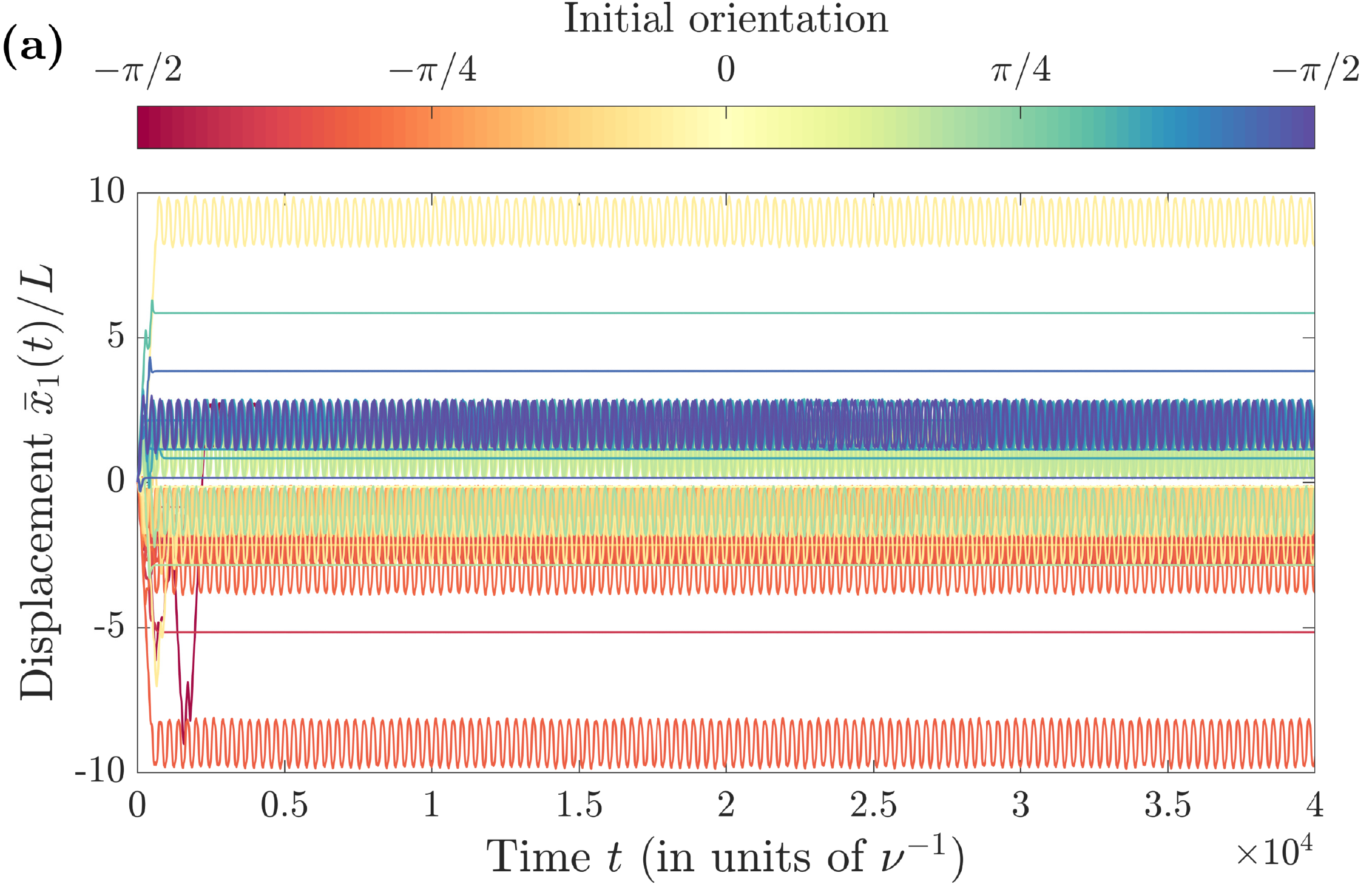}\\[8pt]
  \includegraphics[width=.55\columnwidth]{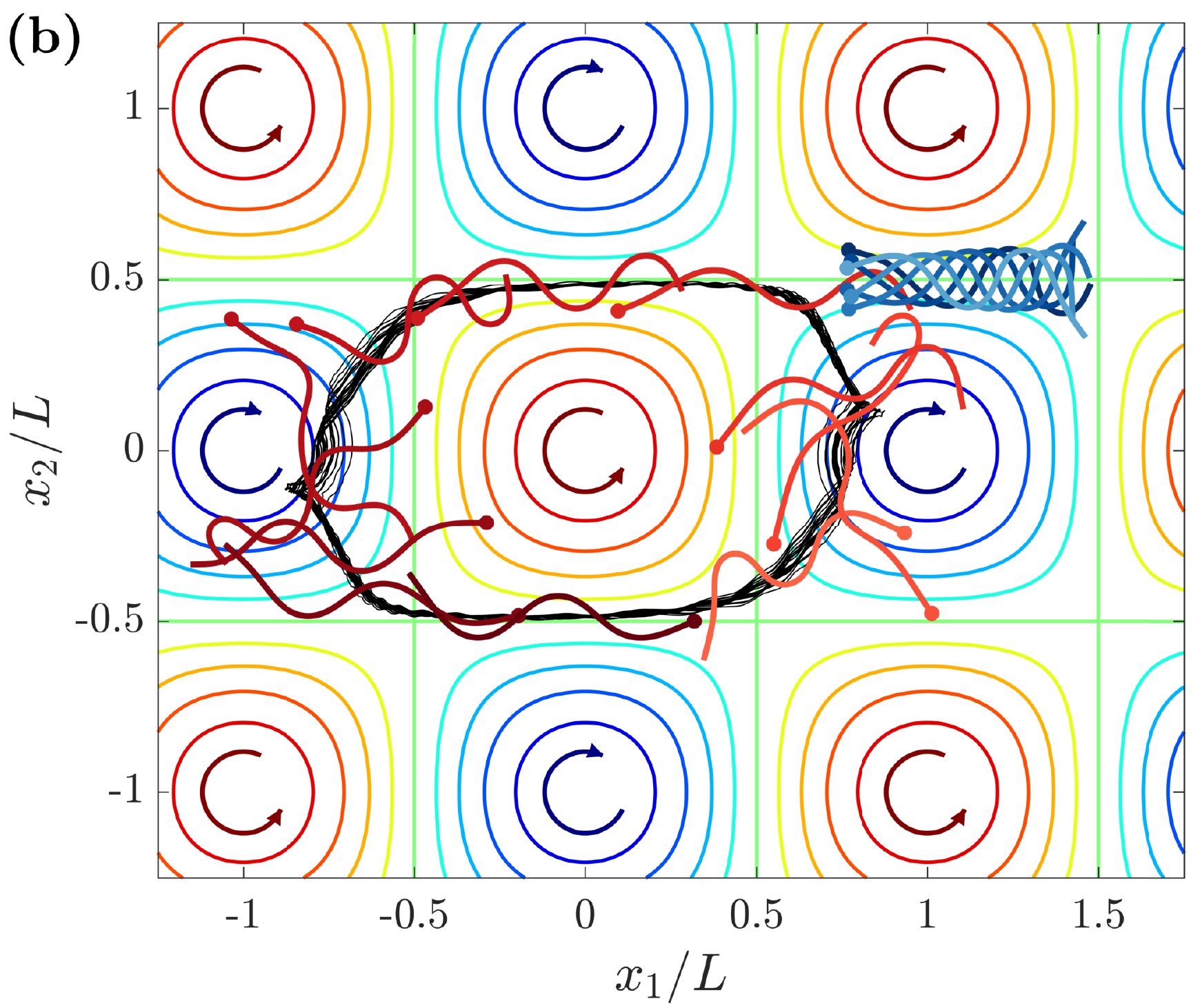}
  \caption{\label{fig_nostrateg} Swimmers continuously undulating in the vertical direction without any specific strategy. The parameters are here $\mathcal{F}=15$, $U = 0.025\ell\nu$ and $\ell/L = 1$. \textbf{(a)}~Displacement along the horizontal direction $x_1$ as a function of time for initially-straight swimmers released with various angles with the $x_1$ axis. \textbf{(b)}~Two instances of trapped swimmers: The blue one is oriented toward $x_1<0$ and is stuck between two cells where it swims against the flow. The red one performs a cycle across several cells, during which it is tumbled back and forth by the flow; The trajectory of the swimmer's centre of mass is shown as a black line. The fluid vorticity $\omega = \partial_1 u_2-\partial_2 u_1$ is represented as coloured contour lines.}
\end{figure}
Figure~\ref{fig_nostrateg}a reports the evolution of the displacement of swimmers initialised with various initial orientations. After crossing a few cells, they get systematically trapped on rather stable cyclic orbits, preventing them from further displacements. We identify two types of cyclic trap, which are illustrated in Fig.~\ref{fig_nostrateg}b. In the case shown in blue, the swimmer is oriented in the wrong direction (towards $x_1<0$) and swims in a counterflow that pushes it to the right and exactly compensates its locomotion. The position of its center of mass barely changes during an undulation period. In the second case, shown in red, the swimmer alternatively swims to the left, is rotated by the flow, swims to the right, then changes again direction, and so on. The mean abscissa $\bar{x}_1(t)$ performs in that case a cyclic motion with an amplitude $\simeq 1.6\,L$ and a period corresponding to approximately 300 forcing periods. The black line shows the position $\bar{\bm X}(t)$ of the swimmer's center of mass sampled over more than 30 cycles. Actually, it does not exactly form a closed loop and tiny deviations can be observed from one cycle to the other. Despite this, such a cyclic motion remains stable and persists for hundreds of cycles. Note that these simulations indicate a very sensitive dependence upon the swimmer's initial orientation as a tiny variation of the initial angle  can lead the swimmer to end up in distant cells of the flow and in different configurations. This sensitivity is a hallmark of a chaotic behaviour. However it also indicates that the swimmers dynamics is not ergodic when they continuously undulate in such a flow. 

Hence the swimmers do not show any net displacement if they just follow their basic swimming procedure without observing any further strategy. Moreover, an adequate navigation policy should be able to prevent, or at least destabilise, the two kinds of trap that were identified. Such an observation can be used to make a guess on adequate minimal observations and actions that should be accounted for in the swimmer's decision process.

\subsection{The optimisation problem}
\label{subsec:optim}
Our problem is to optimise navigation for a swimmer by controlling the parameters of the actuating force based on the current state of the swimmer.  This problem is typically studied using the formalism of \textit{Markov decision processes} (MDPs), which assumes that the \textit{state} of the system is fully observable. This requires grabbing an information that lives, in principle, in the infinite-dimensional set of all two-dimensional curves $s\mapsto\bX(s,t)$ with length $\ell$, and in numerics, in the $(N+1)$-dimensional manifold of $\mathbb{R}^{2N}$ formed by attainable discretised configurations ($N$ being the number of points used to discretise the swimmer arc-length). We hereafter denote by $\mathcal{S}$ this set of states. Because of the high dimensionality of $\mathcal{S}$, a full description of the swimmer state is clearly not possible, neither in numerics, nor in practical applications. 

\begin{figure}[htbp]
\centering
\includegraphics[width=.65\columnwidth]{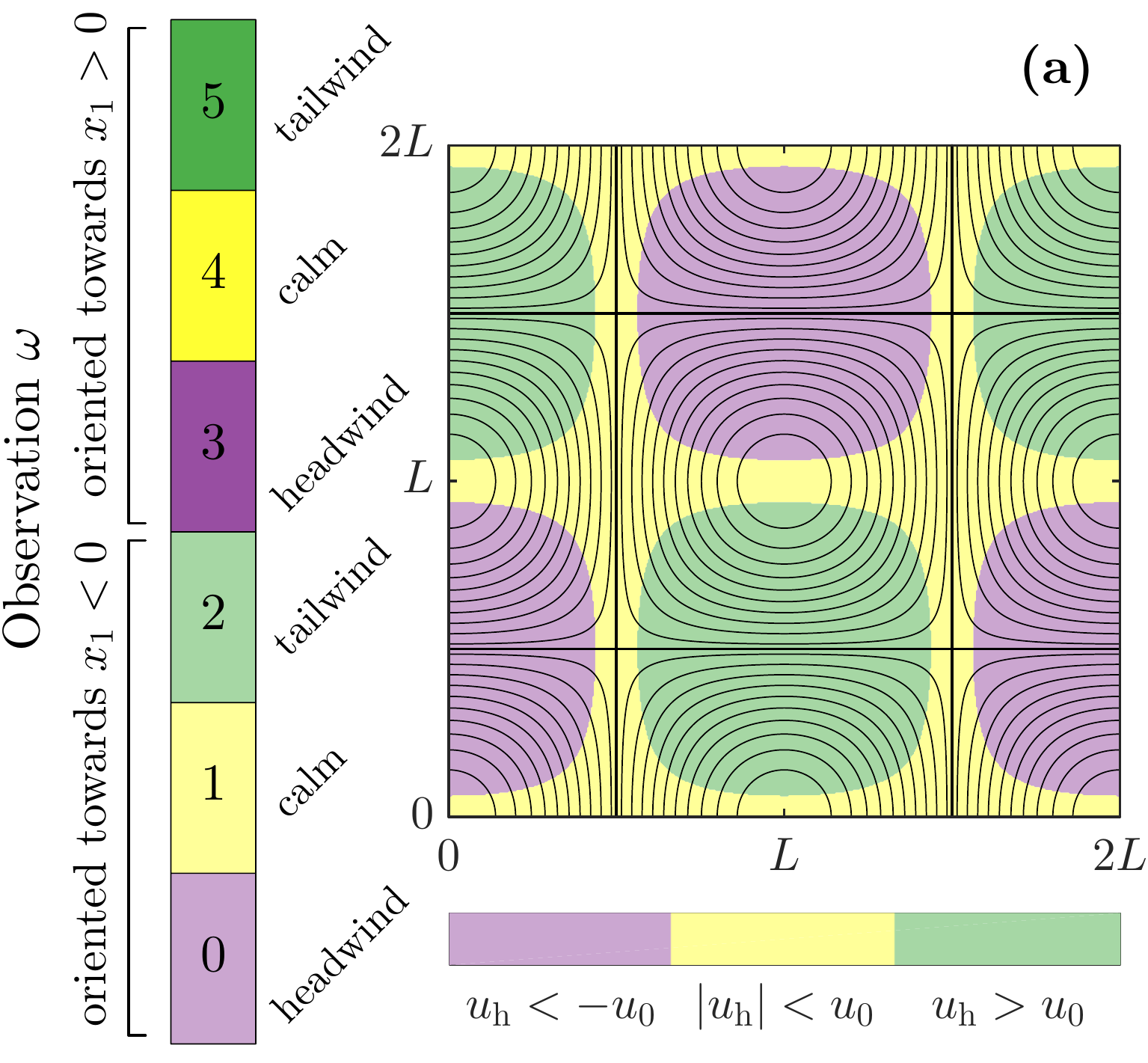}
\includegraphics[width=.65\columnwidth]{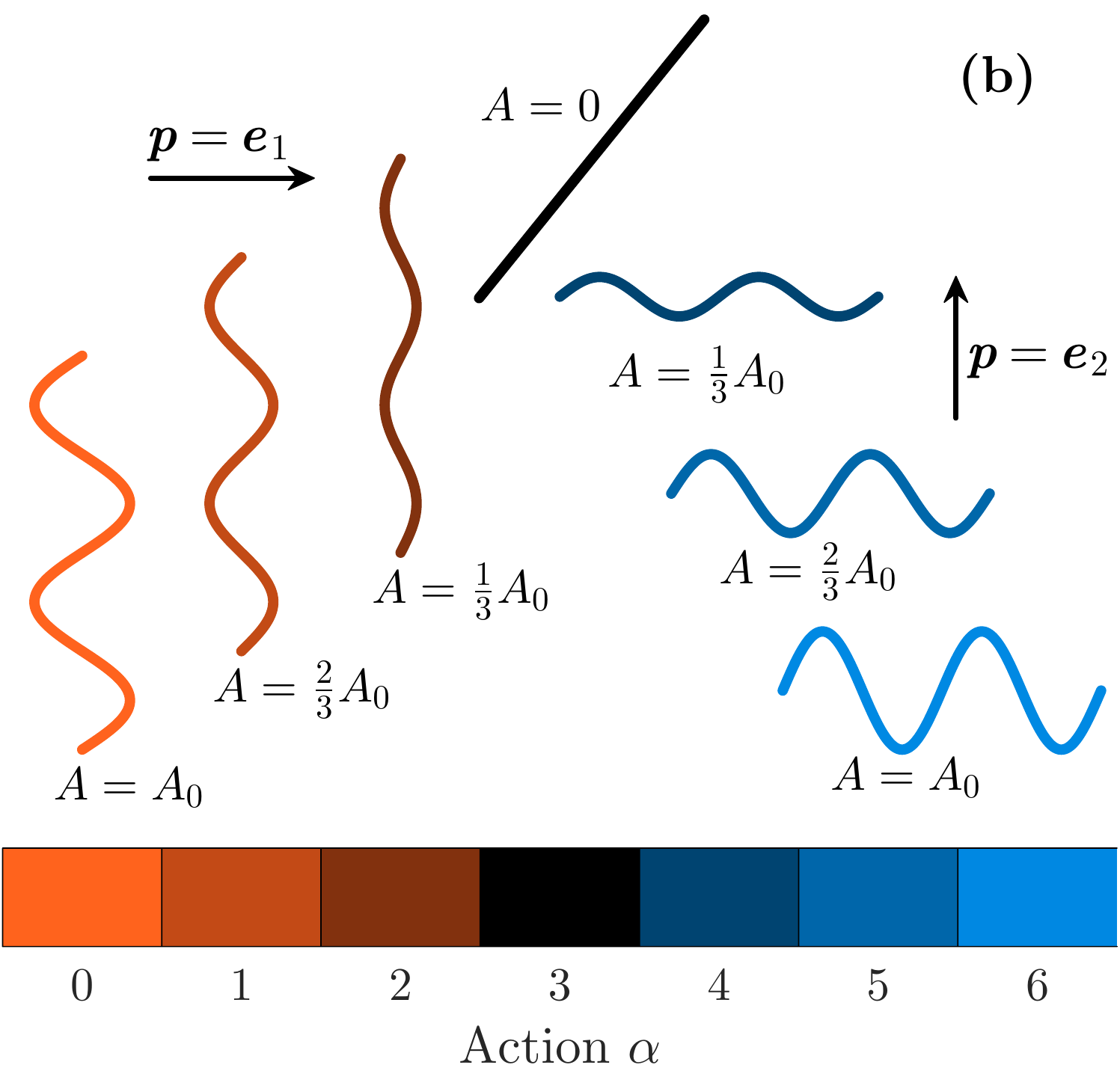}
\caption{\label{fig_sketch} \textbf{(a)} The discretisation of observations depends on both the swimmer orientation, which can be towards positive or negative abscissae, and on the strength of the horizontal fluid velocity at its head $u_{\rm h}$, which divides the flow in regions of three different kinds.  \textbf{(b)} The discretisation of actions set whether the swimmer should propagate an undulation in the horizontal or vertical direction, and with which amplitude $A$.}
\end{figure}
Instead of assuming a full information on the state $\sigma\in\mathcal{S}$, we consider that only a minimalistic information is available. This problem falls under the category of \textit{partially-observable Markov decision processes} (POMDPs), where the observations of the \textit{agent} ---\,the swimmer\,--- are not sufficient to infer the true state of the system. As a result, optimal decision strategies must rely on a limited amount of data, making the problem even more challenging. We denote by $\mathcal{O}$ the set of all possible \textit{observations} $\omega$. We infer from previous section that the swimmer requires information on two features of its state: whether or not it is rightly oriented and whether the fluid velocity helps or hinders its displacement towards $x_1>0$. More specifically, the first property is deduced from the sign of $X_1(0,t) - \bar{x}_1(t)$, namely whether the swimmer's head is located on the right ($\omega = 0,1,2$) or on the left ($\omega = 3,4,5$) of its center of mass. The second property is obtained from the horizontal component $u_{\rm h} = \bm e_1\cdot \bu(\bX(0,t),t)$ of the fluid velocity at the swimmer's head. Three cases are distinguished: either $u_{\rm h}<-u_0$ and the swimmer feels a headwind ($\omega = 0,3$), either $u_{\rm h}>u_0$ and it feels a tailwind ($\omega = 2,5$), or $\vert u_{\rm h}\vert<u_0$ and it feels no significant wind ($\omega = 1,4$). $u_0$ is a parameter that we fix to $u_0/U = 1/5$. This makes up for a total of 6 possible observations that are illustrated and numbered in Fig.~\ref{fig_sketch}(a), so that $\mathcal{O} = \{ 0, 1, 2, 3, 4, 5\}$.

The various \textit{actions} that the swimmer can take are illustrated in Fig.~\ref{fig_sketch}(b). Seven choices are possible, consisting in doing nothing (in black, $\alpha=0$) or applying the active force either in the horizontal ($\bm p = \bm e_1$, in red, $\alpha=0,1,2$) or in the vertical ($\bm p = \bm e_2$, in blue, $\alpha = 4,5,6$) direction, choosing among three possible amplitudes: $A=\frac{1}{3}A_0$ ($\alpha=2,4$), $A=\frac{2}{3}A_0$ ($\alpha=1,5$), or $A=A_0$ ($\alpha=0,6$), where the base non-dimensional amplitude is fixed to $A_0 = 0.08$. The set of all possible actions is again discrete and denoted $\mathcal{A} = \{ 0, 1, 2, 3, 4, 5, 6\}$. 

We assume that the swimmer observes its environment at discrete times $t_n = n \Delta t$ with $n\in\mathbb{N}$. We choose the time step $\Delta t$ smaller than all physical timescales (in practice, we fix $\Delta t = 0.2\,\nu^{-1}$).  A navigation strategy consists in following a \textit{policy} $\pi$, which associates to each couple $(\alpha_n,\omega_n)\in\mathcal{A}\times\mathcal{O}$, a probability $\pi(\alpha_n\vert\omega_n)$ to choose the action $\alpha_n$ having observed $\omega_n$ at time $t_n$. A deterministic policy corresponds to having $\pi(\alpha\vert\omega)=1$ for $\alpha=\alpha_\pi(\omega)$ and $\pi(\alpha\vert\omega)=0$ otherwise. Finding an optimal strategy consists in finding the policy $\pi_\star$ that maximises a given reward over time.

To formally define our POMDP we use the tuple $(\mathcal{S},\mathcal{A},\mathcal{O},{R},{T},\Omega)$, where $\mathcal{S}$, $\mathcal{A}$, and $\mathcal{O}$ are the state, action, and observation sets introduced above. The decision process also depends on the reward function ${R}$,  the transition function ${T}$, and the observation function ${\Omega}$. The reward function ${R}$ maps the current state $\sigma_n\in\mathcal{S}$ and action $\alpha_n\in\mathcal{A}$ to a real number measuring the benefit of having chosen this action. As we are interested in maximising the motion of the swimmer to the right, the chosen reward is the horizontal displacement of its centre of mass ${R} (\sigma_n, \alpha_n) = \bar{x}_1(t_{n+1})- \bar{x}_1(t_n)$.  The transition function ${T}$ is the function that maps the current state and the action taken by the swimmer to the next state: $\sigma_{n+1} = {T} (\sigma_n,\alpha_n)$. Such a function clearly exists because the full dynamics is deterministic and Markovian. Finally, the observation function $\Omega$ is the function that maps the state to the observation sensed by the swimmer: $\omega_n=\Omega(\sigma_n)$. A given policy $\pi$ defines a (possibly stochastic) flow on $\mathcal{S}$: $\sigma_n \mapsto \sigma_{n+1}  = T(\sigma_n,\alpha_n)$ with $\alpha_n$ chosen with probability law $\pi(\cdot\vert\Omega(\sigma_n))$. The policy thus fully determines the sequence $\{(\sigma_n,\alpha_n), n>0\}$ for a given $\sigma_0$.

We aim at finding a policy $\pi$ that maximises the long-term displacement of the swimmer towards positive abscissae. Formalising this optimisation problem requires introducing an adequate objective function.  One could naively think of maximising the actual asymptotic displacement $\lim_{N\to\infty} \left[\bar{x}_1(t_N)-\bar{x}_1(t_0)\right] = \sum_{n=0}^\infty R(\sigma_n, \alpha_n)$. The infinite-horizon sum is however expected to diverge, because we seek policies leading to effective displacement.  Such a pitfall is usually circumvented in MDPs by introducing a discount factor $\gamma$ to ensure convergence. One then maximises the \textit{discounted return}
\begin{equation}
	\mathcal{R}^{\rm disc}[\pi] = \sum_{n=0}^{\infty} {\rm e}^{-\gamma t_n} R(\sigma_n, \alpha_n).
	\label{eq:discount_reward}
\end{equation}
The discount factor $\gamma$ attributes more importance to immediate rewards than to those obtained in a distant future. The choice of this parameter is largely problem-dependent and can have a significant impact on the learned policy.  As seen later, we use such a reward in our implementation of $Q$-learning (Sec.~\ref{subsec:Qlearning}). Still, as discussed in \cite{singh1994learning}, using a discounted reward can be problematic for POMDPs. One can alternatively maximise the so-called \textit{differential return}
\begin{align}
	&\mathcal{R}^{\rm diff}[\pi] = \!\sum_{n=0}^{\infty} \!\left(R(\sigma_n, \alpha_n)-\bar{R}[\pi]\right)\! \nonumber \\
	&\mbox{ where } \bar{R}[\pi] = \!\lim_{N \to \infty} \frac{1}{N}  \sum_{n=0}^N\! \left \langle R(\sigma_n, \alpha_n) \right \rangle\!.
	\label{eq:average_reward}
\end{align}
This formulation weights equally all rewards. It makes use of the mean reward $\bar{R}[\pi]$ that is averaged over both time and the realisations of the POMDP.

In the framework of MDP (for which $\omega\equiv\sigma$), one often introduces the \textit{state value function} $V_{\pi}(\sigma) = \left\langle \mathcal{R}[\pi] \ \vert \ \sigma_0=\sigma\right\rangle$, which quantifies the quality of the policy $\pi$ when we start from the state $\sigma$. A particularly useful function when searching for an optimal policy is the \textit{$Q$-value function}
\begin{equation}
	\mathcal{Q}_\pi(\sigma,\alpha) =  \left\langle \mathcal{R}[\pi] \ \middle\vert \ \sigma_0 = \sigma, \alpha_0 = \alpha \right\rangle.
	\label{eq:defQ_MDP}
\end{equation}
It assesses the value of the policy $\pi$ when taking the specific action $\alpha$ in a given state $\sigma$. Typically, value-based reinforcement learning algorithms try to learn an estimate $\mathcal{Q}_\star$ of the optimal $Q$-value function over all possible policies and use it to extract an optimal deterministic policy $\pi_\star$ as
\begin{align}
  &\pi_\star(\alpha\vert\sigma) = 1 \mbox{ if } \alpha = \alpha_{\pi_\star}(\sigma) = \text{argmax}_{\alpha'} \mathcal{Q}_\star(\sigma, \alpha'), \nonumber\\
  &  \mbox{ and 0 otherwise. }
  \label{eq:opt_policy_MDP}
\end{align}
Such an optimal policy always exists for MDPs, in the sense that it maximises the value function $V_{\pi}(\sigma)$ for all states $\sigma$.

For our partially-observable settings, the agent does not have a full information on $\sigma$ and the the $Q$-value function~(\ref{eq:defQ_MDP}) becomes irrelevant to the navigation problem. A policy that is optimal in the same sense as in MPD is thus no longer guaranteed to exist~\cite{singh1994learning}.   Still, as seen above, one can instead use a different optimality criterion and maximise the differential return~(\ref{eq:average_reward}). Following~\cite{singh1994learning}, the $Q$-value function can be then be defined by projecting $Q_\pi$ on observations, namely
\begin{equation}
	Q_\pi(\omega,\alpha) =   \sum_{\sigma \in \mathcal{S}} \mathcal{Q}_{\pi}(\sigma,\alpha)\,P(\sigma\vert\omega) \ 
\end{equation}
where $P(\sigma \vert \omega)$ is the probability to be in state $\sigma$, given observation $\omega$.

\subsection{A naive strategy}
\label{subsec:naive}

\begin{figure}[b]
  \centerline{\includegraphics[width=\columnwidth]{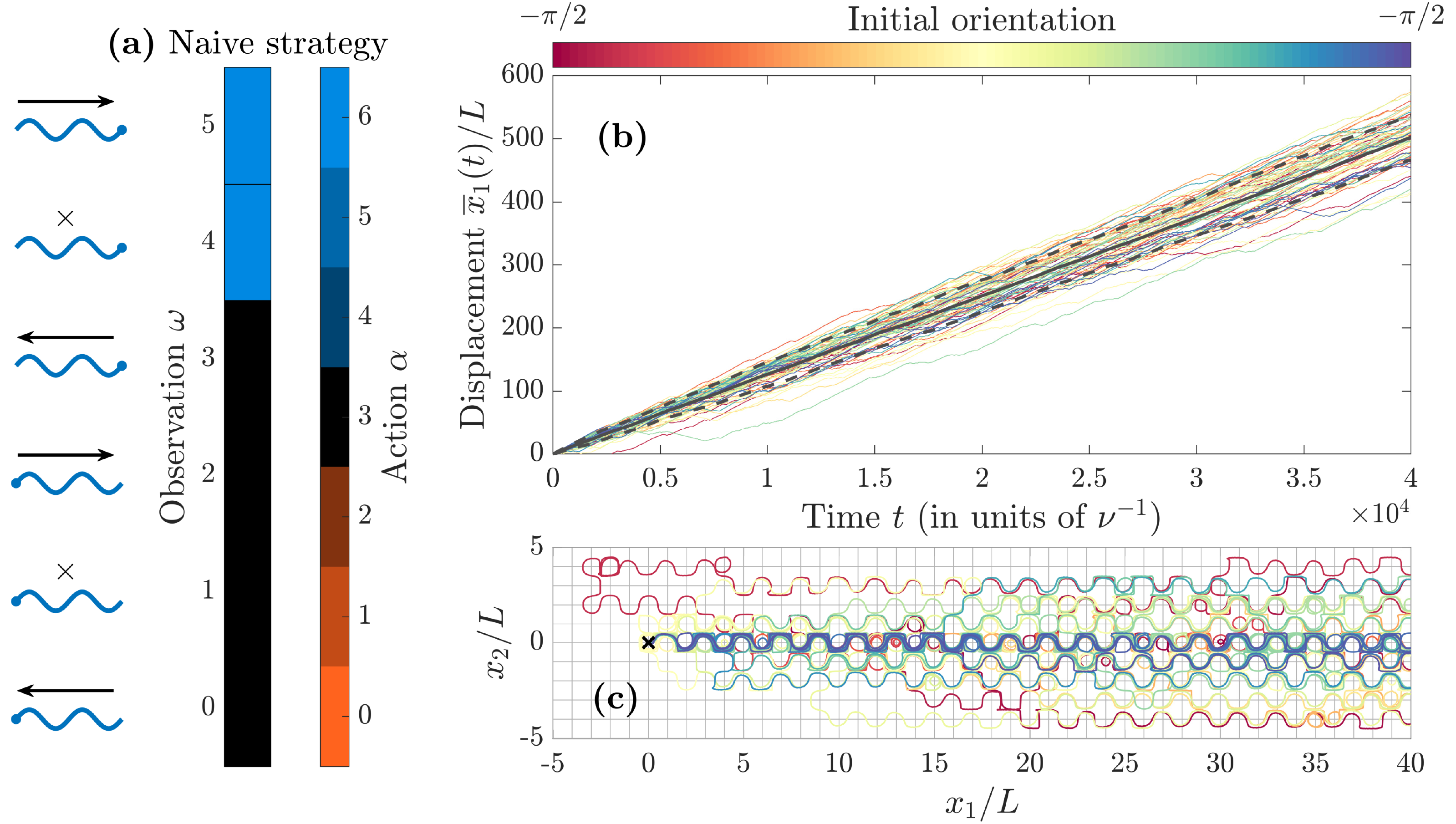}}
  \caption{\label{fig_naive_varIC} \textbf{(a)} Naive strategy, which consists in swimming horizontally with maximum amplitude ($\alpha=6$) whenever the swimmer is rightly oriented and feel a calm or tailwind fluid flow ($\omega=4$ or $5$) and do nothing ($\alpha=3$) otherwise. \textbf{(b)} Horizontal displacement of swimmers initialised with various orientations and following the naive strategy; The average is shown as a bold solid line, and the interval defined by the standard deviation as dashed lines. \textbf{(c)}~Sample of different trajectories in the $(x_1,x_2)$ plane.}
\end{figure}
\begin{figure}[t]
  \centerline{\includegraphics[width=\columnwidth]{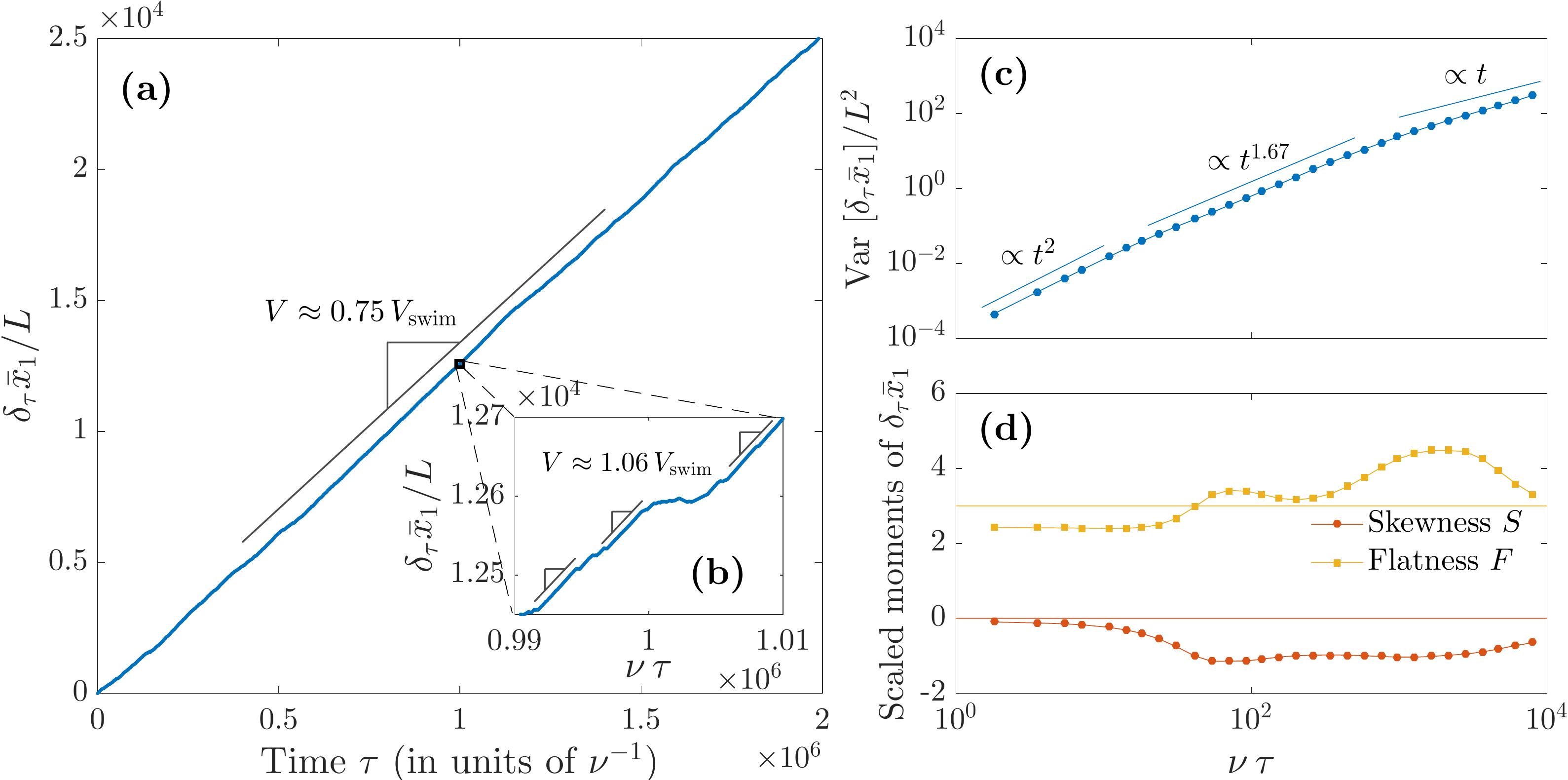}}
\caption{\label{fig_stat_naive} Performance of the naive strategy for $\mathcal{F}=15$, $U=0.025\,\ell\nu$, $\ell/L = 1$, $u_0/U = 1/5$, and $A_0 = 0.08$.  \textbf{(a)} Swimmer's horizontal displacement $\delta_\tau \bar{x}_1 = \bar{x}_1(t+\tau)-\bar{x}_1(t)$, showing an average velocity $V\approx0.75\,V_{\rm swim}$. \textbf{(b)}~Zoom on much shorter times showing a succession of fast displacements and periods of trapping. \textbf{(c)}~Variance of the swimmer's displacement  $\mathrm{Var}\,[\delta_\tau \bar{x}_1] = \langle [\delta_\tau \bar{x}_1-\langle\delta_\tau \bar{x}_1\rangle]^2\rangle$.  \textbf{(d)}~Corresponding skewness  $S = \langle [\delta_\tau \bar{x}_1-\langle\delta_\tau \bar{x}_1\rangle]^3\rangle / (\mathrm{Var}\,[\delta_\tau \bar{x}_1])^{3/2}$ and flatness $F = \langle [\delta_\tau \bar{x}_1-\langle\delta _\tau \bar{x}_1\rangle]^4\rangle / (\mathrm{Var}\,[\delta_\tau \bar{x}_1])^{2}$.}
\end{figure}

We introduce in this section a policy allowing the swimmer to reasonably move in the $x_1>0$ direction. We call it the \textit{naive strategy}. It consists in following rather simple rules: If the swimmer has the proper orientation and simultaneously feels no headwind ($\omega=4,5$), the sinusoidal force (\ref{eq:sinusoidal_force}) is applied with maximal amplitude $A_0$ in the direction $\bm p = \bm e_2$ ($\alpha=6$). If the swimmer is wrongly oriented and faces the $x_1<0$ direction, or experiences a headwind (all other observations), then no force is applied and the locomotion is stopped ($\alpha=3$). This naive policy is shown in Fig.~\ref{fig_stat_naive}a, using a graphical representation that we will employed later on to describe other policies: The different observations $\omega$ are represented as 6 vertically aligned coloured boxes, each colour (from red to blue) standing for the action $\alpha$ taken when $\omega$ is observed. 

This policy breaks the symmetry $x_1\mapsto-x_1$ and thus induces a positive drift. It moreover prevents the swimmer from being indefinitely trapped by similar mechanisms as those observed in Sec.~\ref{subsec:cellflow} in the absence of any strategy. We performed numerical simulations of 100 naive swimmers initialised at $t=0$ at the centre of a cell in a straight configuration, but with different initial orientations spanning $[-{\pi}/{2},{\pi}/{2}]$. As we can see from Fig.~\ref{fig_naive_varIC}b, the naive strategy leads to a positive average displacement, with a distribution of swimmers that perceptibly spreads with time. Sometimes the swimmers temporarily fall in a trap and their displacement stays approximately constant during rather long times.  As seen from the sample of trajectories of Fig.~\ref{fig_naive_varIC}c,  these trapping events correspond to the swimmer turning several times around a given cell before escaping and pursuing its motion towards $x_1>0$.  The quasi-periodic cycles of Fig.~\ref{fig_nostrateg}b are no more stable and the naive strategy makes endless trapping impossible. Thanks to that, all trajectories are asymptotically moving toward $x_1>0$ and the dynamics of swimmers that follow this policy becomes statistically stationary and ergodic.

Figure~\ref{fig_stat_naive} shows more detailed statistics on the displacement of swimmers that follow the naive policy. As can be seen in Fig.~\ref{fig_stat_naive}a, the displacement $\delta_\tau\bar{x}_1 = \bar{x}_1(t+\tau)-\bar{x}_1(t)$ approaches a self-averaged linear behaviour $\delta_\tau\bar{x}_1 \approx V\,\tau$ at large time $\tau$. The average horizontal speed $V$ is approximately $0.75$ times the speed $V_{\rm swim}$ that the swimmer has in the absence of an external flow. When zooming on much shorter timescales (Fig.~\ref{fig_stat_naive}b), one actually observes that this average displacement consists of an alternate sequence of inefficient trapping periods and efficient displacements, during which the swimmer swings smoothly between cells with a speed slightly exceeding $V_{\rm swim}$.  As we will see later, the long-term balance between these two kinds of events is precisely what determines the effectiveness of a given policy.

The variance of $\delta_\tau\bar{x}_1$ is shown in Fig.~\ref{fig_stat_naive}c. Its dependence on $\tau$ follows three successive regimes. At short times $\tau\lesssim 10/\nu$, one has $\mathrm{Var}\,[\delta_\tau \bar{x}_1] \propto \tau^2$, resulting from swimmers moving with an instantaneous velocity different from $V$, and thus deviations $\propto \tau$ from the average displacement. The corresponding higher-order moments of $\delta_\tau\bar{x}_1$ (skewness $S$ and flatness $F$) are shown in Fig.~\ref{fig_stat_naive}d.  One observes at small time lags $S<0$ with $\vert S\vert \ll1$ and thus an almost-symmetric distribution of $\delta_\tau\bar{x}_1$, so that trapping is certainly not contributing much to this regime. Fluctuations are sub-Gaussian, \textit{i.e.}\/ $F<3$.  At larger times, naive swimmers follow an intermediate regime where the variance of $\delta_\tau\bar{x}_1$ grows super-diffusively, approximately as $t^{1.67}$. This regime displays a negative skewness, meaning that trapping is involved. The flatness reaches values above 3, indicating a significant contribution from extreme events. As seen later (Sec.~\ref{subsec:competitive}), this intermediate regime falls in a range during which swimmers have a significant probability to be trapped. It extends to significantly long times, of the order of $\tau\approx500/\nu$, above which the displacement becomes a sequence of independent events. The resulting ultimate regime is diffusive, \textit{i.e.}\/ $\mathrm{Var}\,[\delta_\tau \bar{x}_1] \propto \tau$. The skewness tends asymptotically to $S=0$ and the flatness decreases to possibly approach $F=3$.

We aim at finding policies that outperform this naive strategy. For that, we test in next section various methods of reinforcement learning. It will be important to keep in mind that, even if the swimmer follows a strategy leading to a significant displacement, trapping can be present and result in a significant dependence on history, over times exceeding thousands of undulatory beats.

\section{Reinforcement learning}
\label{sec:reinforcement}

\subsection{$Q$-learning}
\label{subsec:Qlearning}

Here, we first test the performance of classical $Q$-learning. This method, which has been borrowed from MDPs, has been extensively and successfully applied in the past to optimise the navigation of active swimmers~\cite{reddy2016learning,colabrese2017flow,gustavsson2017finding,schneider2019optimal,muinos2021reinforcement,gustavsson2022navigation}.

\subsubsection*{Method}
$Q$-learning is based on the value-iteration update of the Bellman equation. At each step $t_n=n\Delta t$, the swimmer has at disposal an estimation $Q_{t_n}$ of the $Q$-table. It makes an observation $\omega_n$ of its environment, takes an action according to the running policy, which is in the $\varepsilon$-greedy case, is such that $\alpha_n = \textrm{argmax}_{\alpha} Q_{t_n}(\omega_n,\alpha)$ with probability $1-\varepsilon$, other actions being chosen uniformly with probability $\varepsilon/6$.  The swimmer then receives a reward $R_n = \bar{x}_1(t_{n+1}) - \bar{x}_1(t_n)$ and the $Q$-table is updated accordingly. The whole procedure is summarised in Algorithm~\ref{Q-learning}.
\begin{algorithm}[H]
    \caption{$Q$-learning}
    \label{Q-learning}
    Parameters: rates $\lambda$ and $\gamma$; exploration parameter $\varepsilon$
    \begin{algorithmic}[1]
    	\State Initialise $Q$ and $\omega$
    	\For{$n = 1, 2, \dots$}
    	\State Take action $\alpha$  with the $\varepsilon$-greedy law given by $Q(\omega,\cdot)$
    	\State Evolve the swimmer to the new state $\sigma'$
	\State Measure reward $R$ and observation $\omega' = \Omega(\sigma')$
    	\State $Q(\omega,\alpha) \gets (1-\lambda\Delta t)Q(\omega,\alpha)$\\
		\qquad\qquad\qquad\qquad $+ \lambda\Delta t \,[R+{\rm e}^{-\gamma\Delta t}\max_{\alpha'} \!Q(\omega',\alpha')]$
    	\State $\omega \gets \omega'$
    	\EndFor
    \end{algorithmic}
\end{algorithm}

In addition to $\varepsilon\in[0,1]$ that controls how much randomness is put in the learning process, the method depends upon two parameters, which are here appropriately expressed as inverse time scales. The first is the learning rate $\lambda$ that we chose as the inverse of the time needed by a swimmer to cross one cell with velocity $V_{\rm swim}$ in the absence of outer flow, namely here $\lambda=\nu/40$ for $A_0=0.08$.  This rate sets the timescale at which the $Q$-table is updated. A smaller $\lambda$ would have led to adapting the policy with a too long delay compared to the dynamical timescales of the swimmer, and thus to inefficient adjustments. A larger $\lambda$ would imply that the $Q$-table is updated too fast compared to the actual time needed to discern the outcomes of a given action. The second parameter is the discount rate $\gamma$, which sets the horizon of future rewards. It was chosen as the inverse of the time needed by the swimmer to travel across ten cells with $V_{\rm swim}$, namely $\gamma = \nu/400$.  The corresponding timescale is at the edge of the long-correlated regime observed in previous section for the naive policy. Initial entries of the $Q$-table are all set to an arbitrary positive number, equal in our case to $0.25\,L$. 

For MDPs, successive iterations of the procedure~\ref{Q-learning} lead to convergence of the entries of the $Q$-table to the optimal $Q$-value function (\ref{eq:defQ_MDP}) in the limit when $n\to\infty$ and $\varepsilon\to0$ simultaneously. Convergence results rely on the usual stochastic approximation assumptions on the learning rate and are valid as long as all the state-action pairs are indefinitely visited with a positive probability.   The associated empirical greedy policy then converges to the optimal deterministic policy $\pi_\star$ given by~Eq.~(\ref{eq:opt_policy_MDP}). However, such convergence results only hold in the Markovian case. There is no guarantee that they extend to our settings and actually, counter-examples have been constructed in \cite{singh1994learning} showing that $Q$-learning techniques do not generally apply to POMDPs. We nevertheless test this procedure below. 

\subsubsection*{Non-convergence of $\varepsilon$-greedy $Q$-learning}
\begin{figure}[t]
\includegraphics[width=.65\columnwidth]{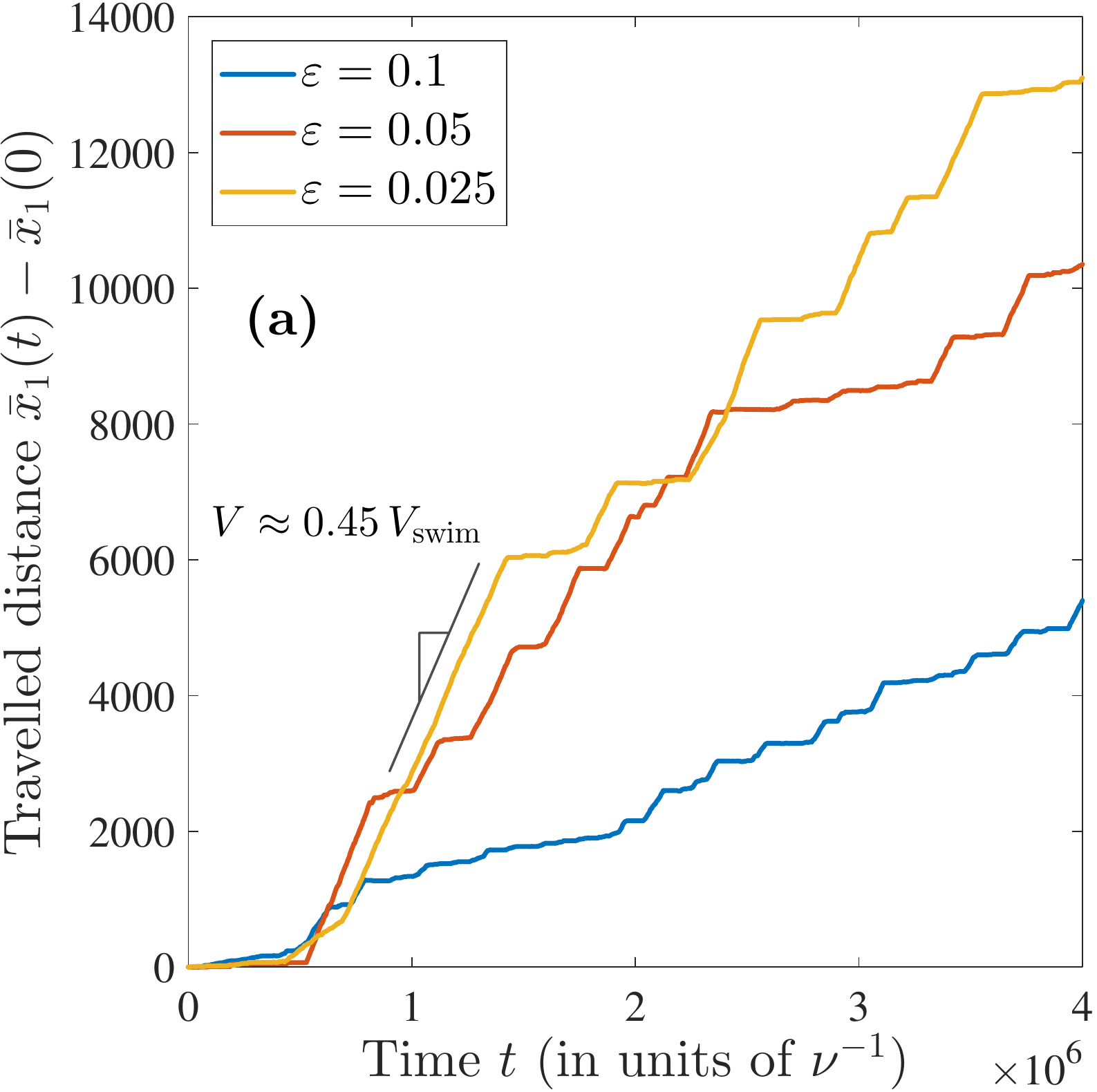}\\[8pt]
\includegraphics[width=.65\columnwidth]{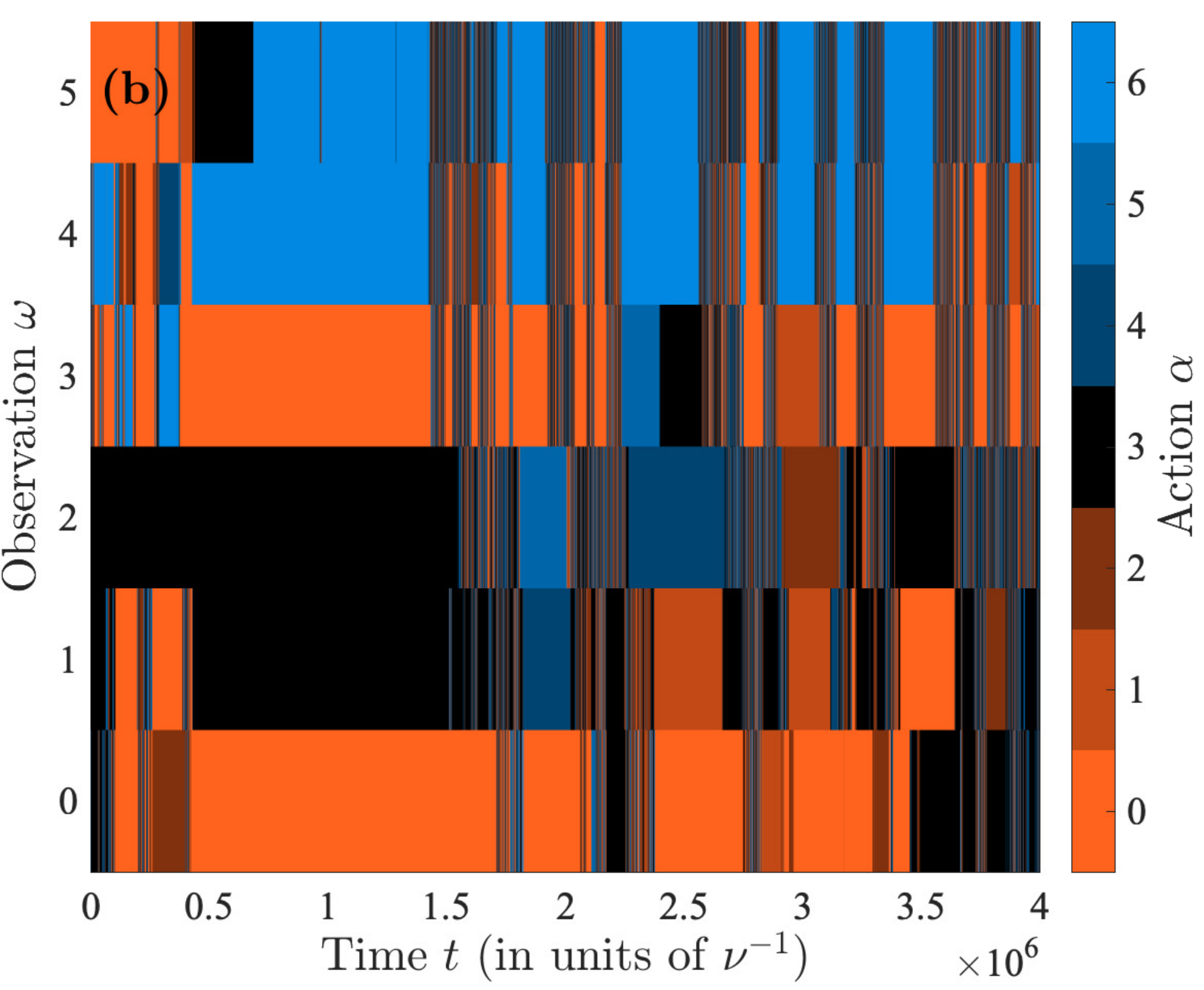}
\caption{\label{fig_QL_var_epsil} Results of $\varepsilon$-greedy $Q$-learning for $\mathcal{F}=15$, $U/(\ell\nu)=0.025$, $\ell/L = 1$, $u_0/U = 1/5$ and $A_0 = 0.08$.  \textbf{(a)} Displacement as a function of time for three different values of the exploration parameter $\varepsilon$. \textbf{(b)} Time evolution of the policy shown here for $\varepsilon = 0.025$.}
\end{figure}

Figure~\ref{fig_QL_var_epsil}(a) shows the displacement of swimmers during the evolution of $Q$-learning for decreasing values of the exploration parameter $\varepsilon$. All instances lead to a net displacement of the swimmer. It consists of long periods of forward motions interrupted by phases during which the swimmer barely progresses. These alternations become less and less frequent when $\varepsilon$  decreases. Figure~\ref{fig_QL_var_epsil}(b) shows the time-evolution of the policy followed by the swimmer for $\varepsilon=0.025$. Each extended period of forward motion corresponds to a stabilisation of the running policy. For instance, between times $t=0.7$ and $1.4\times10^6 \nu^{-1}$, the swimmer maintains an average horizontal velocity $\approx 0.45\,V_{\rm swim}$ that is smaller, but comparable to the performance of the naive strategy. During this time interval, the swimmer follows a policy that differs from the naive one only by favouring a vigorous horizontal undulation ($\alpha = 0$, bright red) when a headwind is observed ($\omega = 0$ and $3$). This temporarily learned policy is however forgotten at times $t>1.5\times10^6 \nu^{-1}$. Other sustainable strategies are selected later on, giving rise to subsequent periods of forward motion with different, but comparable horizontal velocities. These numerical experiments obtained at varying $\varepsilon$ allow us to extrapolate to what would be obtained if the level of randomness were decreased progressively: As the duration of forward-motion periods expands when $\varepsilon$ increases, the learning procedure will probably get stacked to a given policy determined by the history of the swimmer's trajectory and thus very unlikely to be the optimum. This gives evidence that $Q$-learning methods do not easily converge for our problem. 

\begin{figure}[h]
\centering
\includegraphics[width=\columnwidth]{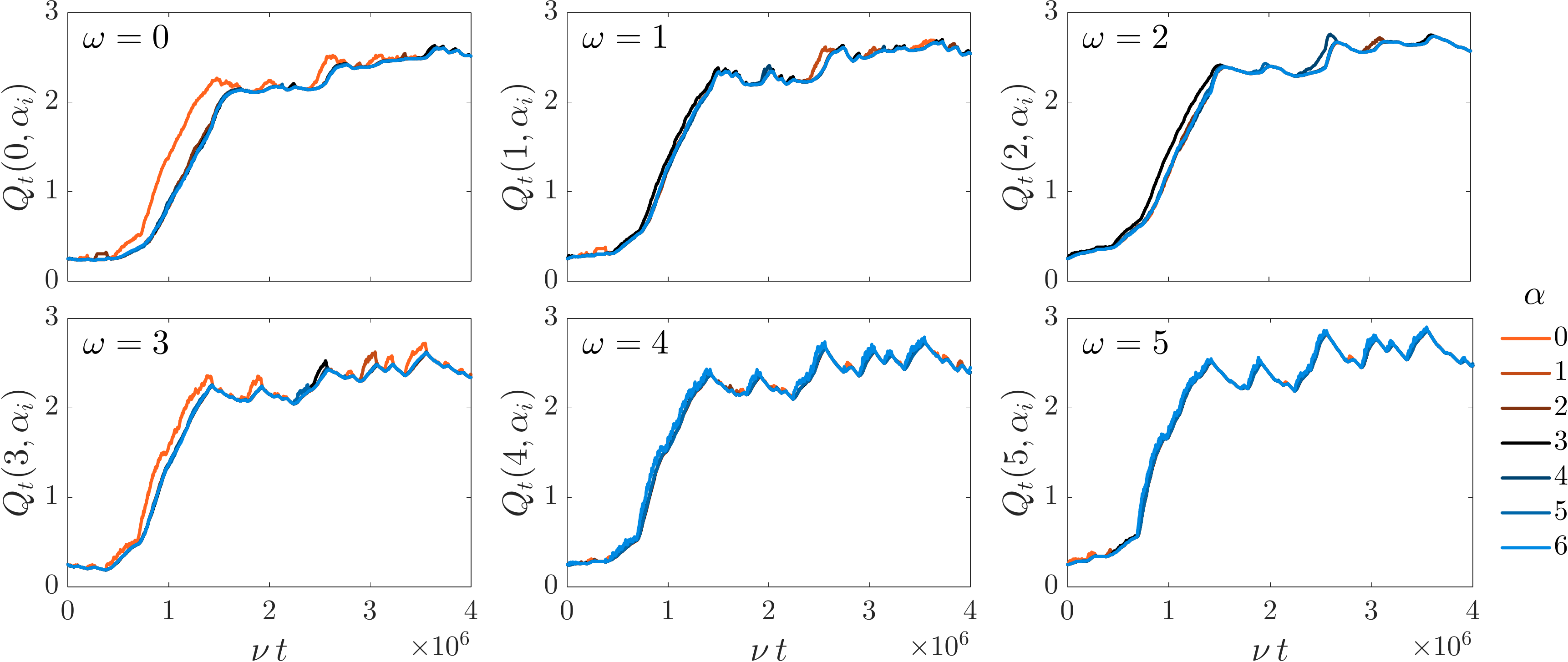}
\caption{\label{fig_QL_var_epsil_convergQ} Time evolution of the different components of the $Q$ table obtained for $\varepsilon = 0.025$, as in Fig.~\ref{fig_QL_var_epsil}b. The six panels correspond to the various values of the observation $\omega$, while the different colours stand for the action $\alpha$, as labeled.}
\end{figure}
We interpret the above-mentioned loss of memory as a consequence of long-term trapping phases that can clearly not be detected from our reduced set of observations.  The underlying mechanism gets clearer when looking at the time evolution of the $Q$-table entries in Fig.~\ref{fig_QL_var_epsil_convergQ}. The periods of forward motion are associated with an increase of all $Q_t(\omega,\alpha)$, with the current running policy weakly singling out given entries. Once the swimmer enters a phase of quasi immobilisation, this growth stops and all entries of the $Q$-table decrease simultaneously, without any possibility to keep in mind the previously learned strategy.  Hence, convergence could in principle be only achieved if the learning rates is small enough to filter out such trapping events, and would thus require running the $Q$-learning algorithm for extravagantly long times.

\subsubsection*{An iterative Markovian approximation}
Motivated by the suspicion that convergence could require very long times, we test here the idea to approximate the dynamical evolution of the swimmer by an MDP. Our hope is that this approximation will capture the most relevant information of our optimisation problem, namely, the transition probabilities between the states of our environment and the distribution of the rewards obtained by our agent. The advantages of this approach are twofold: First, since the MDP only cares about the transitions and the rewards process abstracting away all the other aspects of the dynamics, the associated learning algorithms will run significantly faster, without having to simulate simultaneously the whole swimmer dynamics; Second, this approach would separate the issue of non-Markovianity from other potential difficulties.

Our procedure consists in constructing a sequence of policies $\pi_0$, $\pi_1$, \dots $\pi_k$ that will hopefully converge to the optimal $\pi_\star$. At each step, we simulate a swimmer that follows the policy $\pi_k$, trying out at every time step $t=t_n$, all possible actions to monitor the new observation and reward at time $t_{n+1}$. This is used to construct numerical approximations to the transition probability $p_{{\rm T},k}(\omega'\vert\omega,\alpha)$ of observing $\omega'$ at time $t+\Delta t$ given that $\omega$ was observed and action $\alpha$ was performed at time $t$, together with the corresponding distribution of rewards $p_{{\rm R},k}(R\vert\omega,\alpha)$. Both distributions depend of course on $\pi_k$. We then use the approximate probabilities $p_{{\rm T},k}$ and $p_{{\rm R},k}$ to run the $Q$-learning algorithm that, because of the Markovian formulation imposed now, is ensured to converge.  This leads to construct the optimal policy $\pi_{k+1}$ associated to the approximate system. This procedure is reiterated changing the base policy to $\pi_{k+1}$, until it attains a fixed point. The method is summarised in Algorithm~\ref{iterative_Markov}.
\begin{algorithm}[H]
    \caption{Iterative Markovian approximation}
    \label{iterative_Markov}
    \begin{algorithmic}[1]
    	\State Initialise policy $\pi_0$
    	\Repeat \text{ } for $k=0,1,2, \ldots$
	\State Simulate a swimmer that follows policy $\pi_k$
	\State Measure $p_{{\rm T},k}(\omega'\vert\omega,\alpha)$ and $p_{{\rm R},k}(R\vert\omega,\alpha)$
    	\State Use them to find the optimal policy $\pi_{k+1}$
    	\Until{$\pi_{k+1} \in \{ \pi_0, \pi_1, \ldots, \pi_k\}$}
    \end{algorithmic}
\end{algorithm}

The motivation behind this procedure is that, if the Markovian approximation is not too far off, then it is natural to think that the optimal policy $\pi_{k+1}$ of the approximate system should be at least an improvement on the policy $\pi_k$ if not also the optimal policy when we go back to the real system. Hence, if the optimal policy $\pi_\star$ is a fixed point of our procedure, then the sequence $\{\pi_k; k \geq 0\}$ would converge to it, thus solving our problem.

We have run this procedure, choosing for the initial policy $\pi_0$ the naive strategy of Sec.~\ref{subsec:naive}.  After three iterations the algorithm circled back to the policy we encountered on the first iteration $\pi_3 = \pi_1$. Hence this proposed procedure does not lead to any improvement with respect to the naive policy. This could be again a sign of the highly non-Markovian nature of our setting. We therefore test in the next section various approximation-based methods that could in principle lead to efficient results for POMDPs.

\subsection{Approximation-based methods}

In the previous section, we made use of the traditional $Q$-learning with discounted return~(\ref{eq:discount_reward}) to estimate the action-value function. We applied blindly this method by replacing states with observations and obtained only limited success. Here, we will explore two approaches that belong to the broad class of \textit{approximation methods} for reinforcement learning~\cite{sutton2018reinforcement}: the \textit{semi-gradient differential SARSA} and the \textit{Actor-Critic policy-gradient} method. Both use a formulation of the optimisation problem in which value functions are estimated in terms of the differential return~(\ref{eq:average_reward}) instead of the discounted return.

The main motivation for using such approximation methods is the partially-observable nature of our problem. In such settings, accurate estimations of the action-value function $Q$ are difficult, hindering the convergence of exact-solution algorithms like $Q$-learning~\cite{singh1994learning}. However, by using approximation methods, such as neural networks or other parametric models, we can represent the policy and value function in a way that takes into account only the available observations rather than the full state. Such methods are flexible and effective and, in particular, they provide a way to trade-off between the quality of the solution and computational complexity. This makes them a good choice for problems with large or continuous state spaces, where exact solution methods are not applicable. They can also search for optimal stochastic policies, which can help ensure exploration during the learning process, particularly when the optimal policy may not be deterministic, as is often the case in POMDPs~\cite{singh1994learning}, though not likely in our exact case. For these reasons, approximation methods allow us to effectively address the partial observability issue and achieve good performance, at least in theory, without compromising the underlying theory of reinforcement learning.

\subsubsection{Semi-gradient differential SARSA}

The semi-gradient differential SARSA algorithm is a value-based method, like $Q$-learning.  It similarly builds on the idea of estimating the action-value function $Q$ to construct an optimal  policy, but uses for that the differential return instead of the discounted return.  A key difference between this method and traditional SARSA or $Q$-learning is that it involves an approximation of the $Q$-function in place of its exact value. We use here the linear parametrisation 
$\mathcal{Q}(\sigma,\alpha) \approx \hat{\mathcal{Q}}_{\boldsymbol{\eta}}(\sigma,\alpha) = \sum_{ij} \eta_{ij}\,\delta_{\Omega(\sigma),i}\,\delta_{\alpha,j}$ 
where $\delta$ is the Kronecker delta, $\sigma \mapsto \Omega(\sigma) = \omega$ is the 
observation function introduced in Sec.~\ref{subsec:optim}, and $\boldsymbol{\eta}\in\mathbb{R}^6\times\mathbb{R}^7$ denotes the approximation parameters.  
This approach aggregates together all states leading to 
the same observation (similarly to what we did for $Q$-learning). 
The partial observability of the problem is then devolved to this specific choice of the approximation. Such a method was used successfully in~\cite{berti2022reinforcement} to find the optimal swimming strategy for Najafi's swimmer~\cite{najafi2004simple}

The main idea of semi-gradient differential SARSA is to update the approximation of the value function
by combining the gradient descent and temporal difference (TD) learning methods in order to converge to the parameters 
$\boldsymbol{\eta^*}$ that approximate best the optimal $Q$-function. 
The action at the $n$-th step is chosen, as in $Q$-learning, 
such that $\alpha_n = \mathrm{argmax}_\alpha \hat{\mathcal{Q}}_{\boldsymbol{\eta}}(\sigma_n,\alpha)$ 
with possibly an $\varepsilon$-greedy step. The resulting procedure is summarised 
in Algorithm~\ref{alg_SARSA}.
\begin{algorithm}[H]
    \caption{Semi-gradient differential SARSA}
    \label{alg_SARSA}
    Algorithm parameters: rates $\lambda_1, \lambda_2$; exploration parameter $\varepsilon$
    \begin{algorithmic}[1]
    	\State Initialise $\omega$, $\alpha$, $\bar{R}$, and the approximation parameters $\boldsymbol{\eta}$
    	\For{$n = 1, 2, \dots$}
    	\State Take action $\alpha$ and evolve to the new state $\sigma'$
    	\State Measure reward $R$ and observation $\omega' = \Omega(\sigma')$
    	\State Choose action $\alpha'$ with $\varepsilon$-greedy law given by $\hat{\mathcal{Q}}_{\boldsymbol{\eta}}(\sigma',\cdot)$
    	\State Compute the error $\delta = R - \bar{R}+ \hat{\mathcal{Q}}_{\boldsymbol{\eta}}(\sigma',\alpha') - \hat{\mathcal{Q}}_{\boldsymbol{\eta}}(\sigma,\alpha)$
    	\State $\bar{R} \gets \bar{R} + \lambda_1\Delta t\,\delta$
    	\State $\boldsymbol{\eta} \gets \boldsymbol{\eta} + \lambda_2\Delta t\,\delta\,\nabla\hat{\mathcal{Q}}_{\boldsymbol{\eta}}(\sigma,\alpha)$
    	\State $\omega \gets \omega'$, $\alpha \gets \alpha'$
    	\EndFor
    \end{algorithmic}
\end{algorithm}

\begin{figure}[h]
  \includegraphics[width=.65\columnwidth]{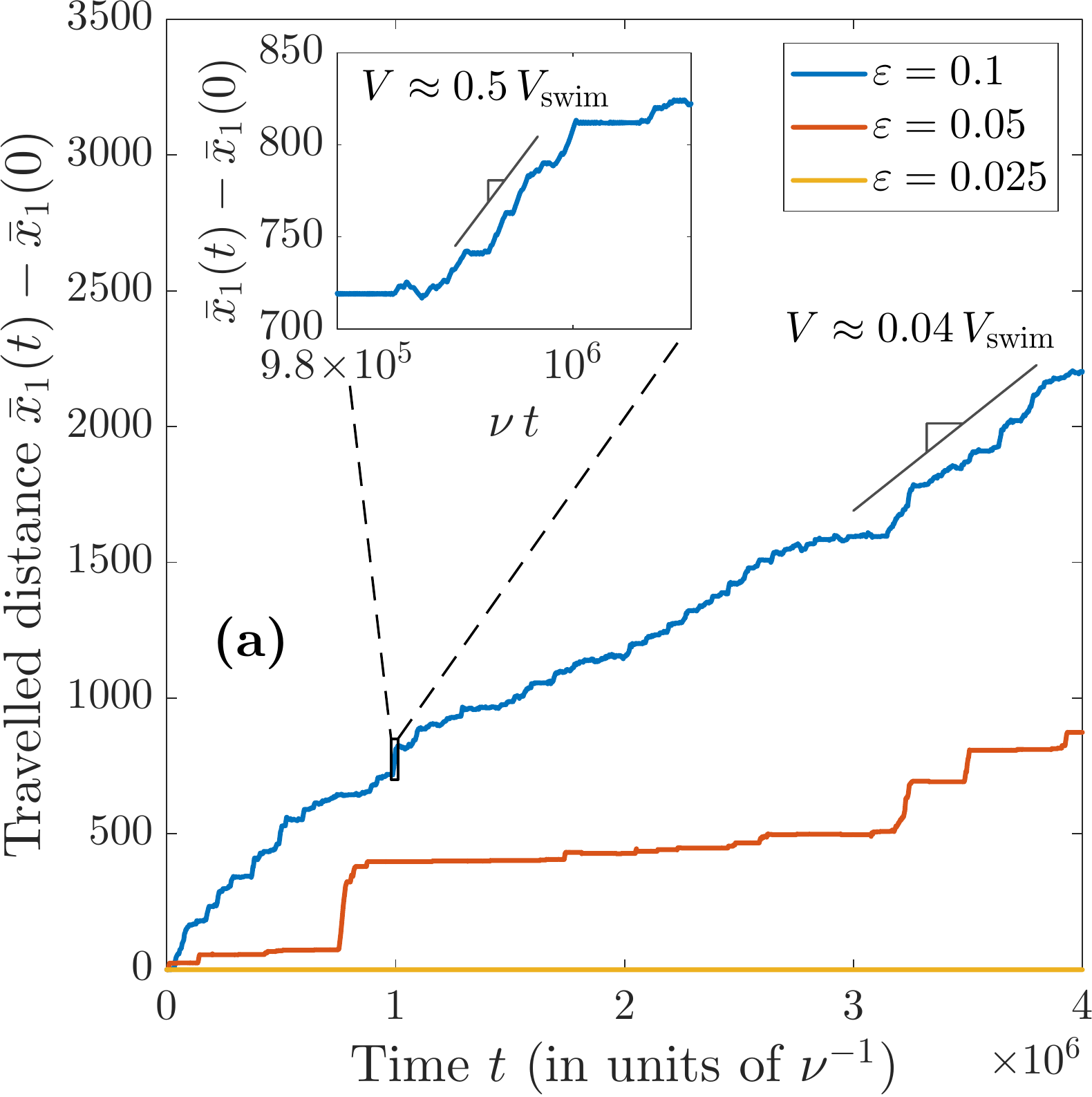}\\[8pt]
  \includegraphics[width=.65\columnwidth]{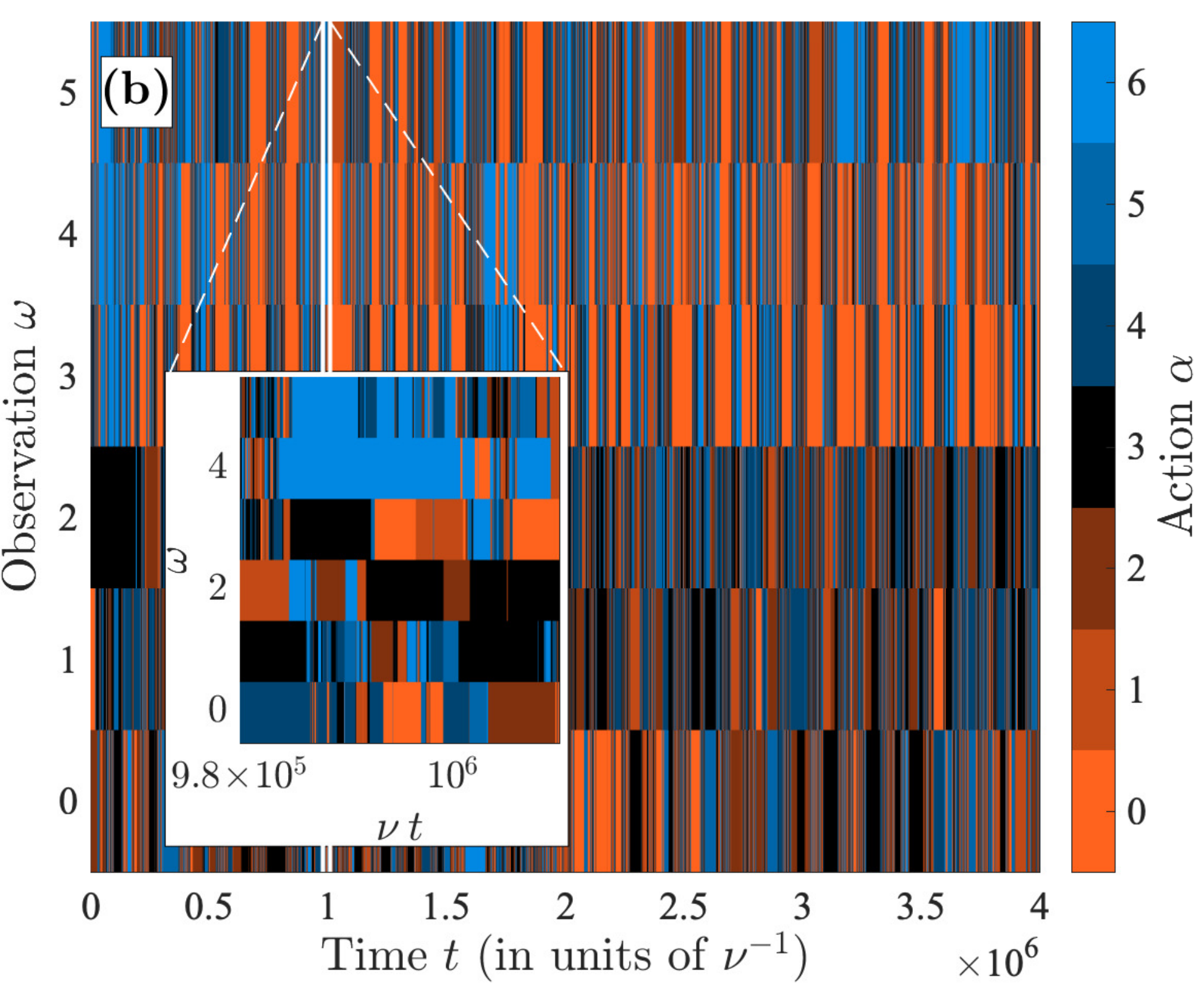}
  \caption{\label{fig_SARSA_var_epsil} Results of semi-gradient differential SARSA obtained with the same physical parameters as in Fig.~\ref{fig_QL_var_epsil} of previous subsection.  \textbf{(a)}~Time-evolution of the displacement for three different values of the exploration parameter $\varepsilon$. \textbf{(b)} Time evolution of the optimal policy shown here for $\varepsilon = 0.1$. Both figures show as insets a zoom on a time interval during which the swimmer significantly progresses.}
\end{figure}

Figure~\ref{fig_SARSA_var_epsil} reports results obtained with this method. We have here chosen the rates $\lambda_1 = 0.025\nu$ and $\lambda_2=0.025\nu$, which corresponds to the inverse of the times needed by the swimmer to cross one and ten cells with $V_{\rm swim}$, respectively. The displacements obtained for different values of the exploration parameter $\varepsilon$ (Fig.~\ref{fig_SARSA_var_epsil}a) are by an order of magnitude smaller than those resulting from the $\varepsilon$-greedy $Q$-learning algorithm. In addition, results indicate that the learning performance decreases when $\varepsilon$ decreases, at variance with what was observed for $Q$-learning. At the largest value of $\varepsilon$, one finds forward-moving periods to be much shorter and trapping phases much more frequent. Still, when zooming on an interval of time when the swimmer significantly progresses, one observes that the local velocity is comparable to those obtained with $Q$-learning. As can be seen from Fig.~\ref{fig_SARSA_var_epsil}b, the corresponding running policy fluctuates much, and significant displacement only occurs when the policy is able to maintain for a significant amount of time the action $\alpha=6$ (undulate vertically with maximal amplitude) for the two most favourable observations $\omega=4$ and $5$ (corresponding to being rightly oriented with no headwind).

We found that in our setting, the semi-gradient differential SARSA method is not able to learn properly
due to a non-ergodicity of the environment. Indeed, the swimmer is often trapped in a situation where 
it observes the same observation $\omega=3$ (being wrongly oriented with no headwind), performs the same 
action $\alpha=6$ (undulate vertically with maximal amplitude), and remains indefinitely trapped in this situation.
(The curve associated to $\varepsilon=0.025$ in Fig.~\ref{fig_SARSA_var_epsil}a is an example of such a situation.)
This is due to the fact that in a large set of configurations leading to observation $\omega=3$, the swimmer
performing action $\alpha=6$ remains in the same set of configurations.  Furthermore, the swimmer keeps on being stuck, even if it performs other actions but not for a long-enough time, so that its probability of escaping decreases exponentially fast as $\varepsilon\to0$.

\subsubsection{Actor-Critic policy-gradient method}

Policy-gradient methods strongly differ from $Q$-learning and semi-gradient differential 
SARSA in that, instead of learning the function $Q$, they learn directly the policy $\pi$ 
by interacting with the environment. Additionally, instead of using the temporal difference rule 
to learn the estimates, policy-gradient methods are gradient-based, meaning that the policy is 
itself approximated, similarly to the value function for differential SARSA, by an estimate $\pit$, which involves a set of parameters $\boldsymbol{\theta}$ that are learned using gradient descent.

We furthermore use here an \textit{Actor-Critic} version of such a method. The ``actor'' represents 
the policy, while the ``critic'' estimates the value function. This separation can help improve the 
stability and convergence of the policy-gradient algorithm, as well as reduce the variance of the 
gradient samples used to update the policy parameters.
Together, the actor and the critic form a coalition where the actor selects actions and the critic 
evaluates the quality of those actions to provide feedback to the actor to improve its policy.

\begin{figure}[t]
\centering
\includegraphics[width=0.325\columnwidth]{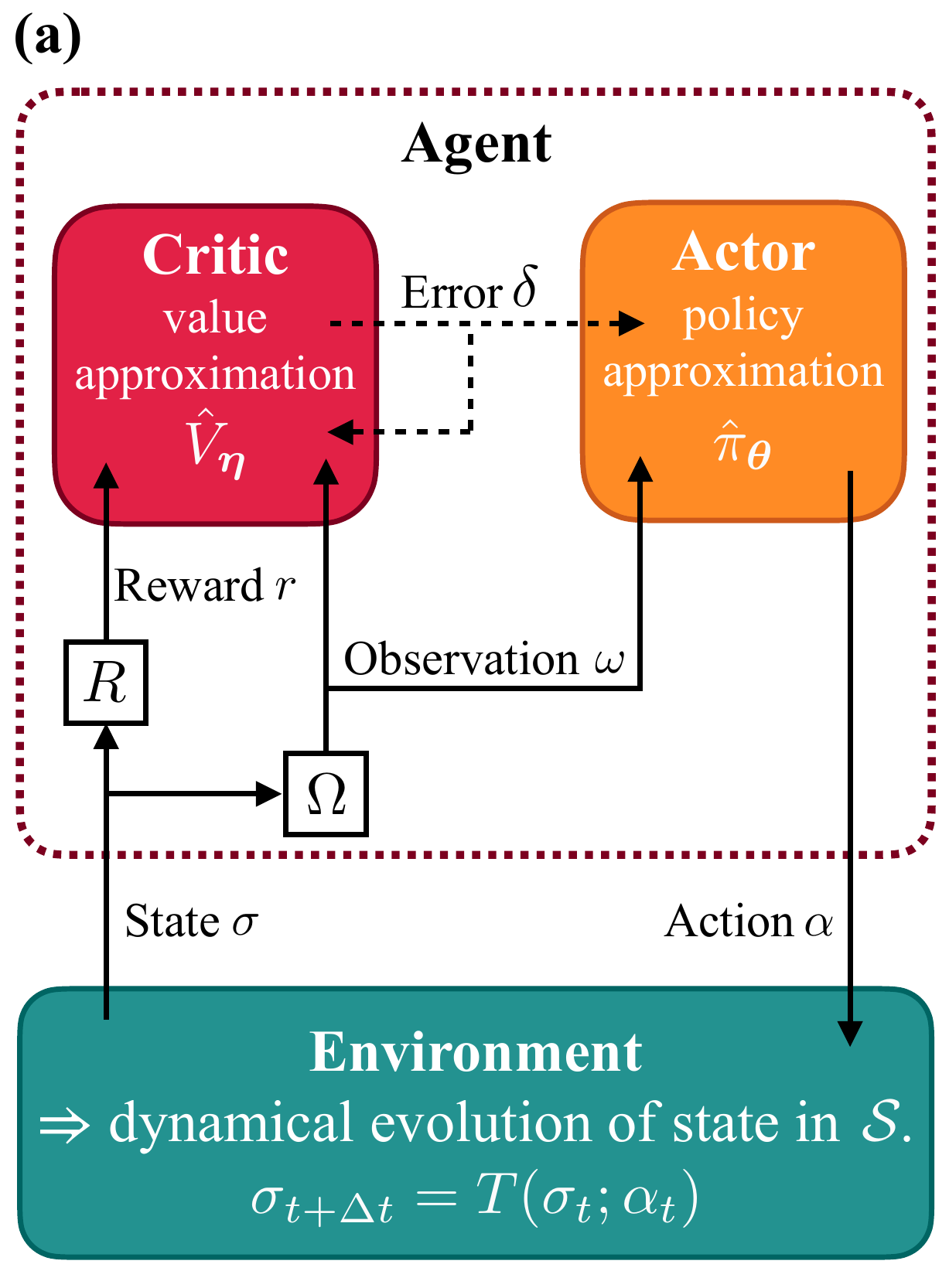}
\hfill
\includegraphics[width=0.65\columnwidth]{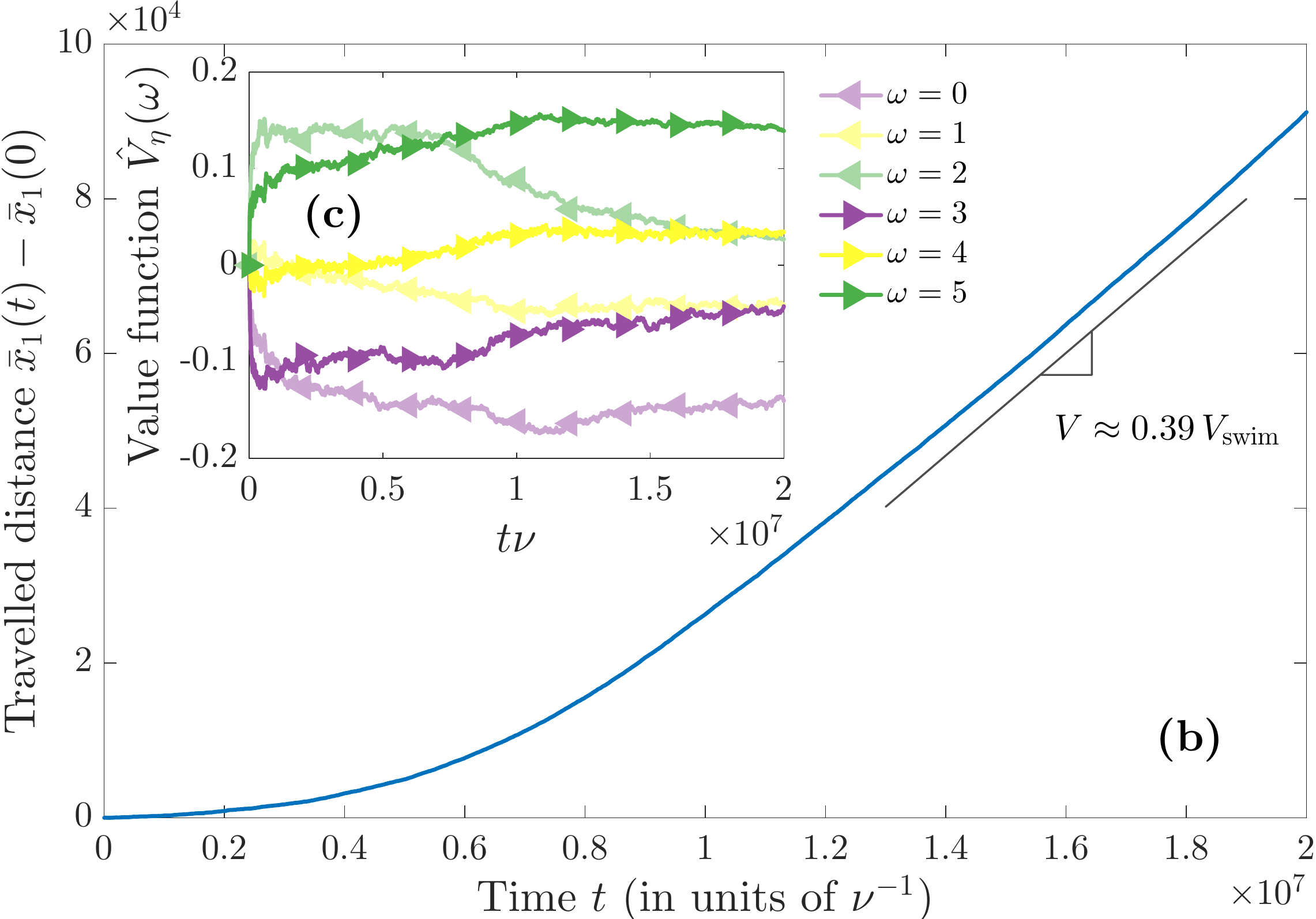}
\caption{\label{fig_actor_critic_displ} \textbf{(a)} Sketch of the actor-critic algorithm. \textbf{(b)} Time evolution of the displacement obtained for the actor-critic algorithm for fixed hyper-parameters. Inset \textbf{(c)} Time evolution of the value function for the six different values of the observation $\omega$, as labelled.}
\end{figure}

The general scheme of the Actor-Critic algorithm is sketched in Fig.~\ref{fig_actor_critic_displ}a. After a change in the environment, both the actor and the critic are informed about the new observation $\omega$ of the system. The critic, which has also access to the reward, updates its approximation $\hat{V}_\eta$ of the value function and communicates to the actor the temporal-difference (TD) error $\delta$, which measures the difference between the expected return and the actual return.  The actor uses the information that $\delta$ provides on the quality of the approximated policy $\pit$ in order to update it and decides, according to the observation $\omega$,  the action to be taken throughout the next step.

We choose to represent the policy by the soft-max parametrisation
\begin{equation}
	\pi(\alpha\vert\sigma) \approx \pit(\alpha\vert\sigma) = \sum_{ij} \frac{1}{\mathcal{Z}_i} {\rm e}^{\theta_{ij}}\delta_{\Omega(\sigma),i}\,\delta_{\alpha,j},
	\label{eq:soft-max}
\end{equation}
with normalising factor  $\mathcal{Z}_i = \sum_j {\rm e}^{\theta_{ij}}$. The approximated policy hence depends on the state $\sigma$ only through the observation $\omega = \Omega(\sigma)$. This seamlessly takes into account partial observability by considering only the available information rather than the full state of the system.  The policy parameters $\boldsymbol{\theta}\in\mathbb{R}^6\times\mathbb{R}^7$ are optimising a \textit{performance measure} given by the average return $\bar{R}[\pi_{\boldsymbol{\theta}}]$ defined in Eq.~(\ref{eq:average_reward}).  The gradient-ascent procedure used by the actor to update $\boldsymbol{\theta}$ requires to approximate the gradient $\nabla_{\boldsymbol{\theta}} \bar{R}[\pi_{\boldsymbol{\theta}}]$ of the performance measure. We rely on the policy-gradient theorem (see, \textit{e.g.}, \cite{sutton2018reinforcement})
\begin{align}
	\nabla_{\boldsymbol{\theta}} \bar{R}[\pi_{\boldsymbol{\theta}}] &= \left\langle \mathcal{Q}_{\pit}(\sigma, \alpha) \frac{\nabla_{\boldsymbol{\theta}} \pit (\alpha \vert \sigma)}{\pit (\alpha \vert \sigma)} \right\rangle \nonumber\\
	&= \left\langle \mathcal{Q}_{\pit}(\sigma, \alpha) \, \nabla_{\boldsymbol{\theta}} \log \pit (\alpha\vert\sigma) \right\rangle,
\end{align}
which allows us to instantiate the performance-measure gradient as ${\nabla}_{\boldsymbol{\theta}} \bar{R}[\pi_{\boldsymbol{\theta}}] \approx \hat{\mathcal{Q}}_{\boldsymbol{\eta}}(\sigma, \alpha) \, \nabla_{\boldsymbol{\theta}} \log \pit(\alpha\vert\sigma)$,
where $\hat{\mathcal{Q}}_{\boldsymbol{\eta}}$ is an approximation of the value function, at the hands of the critic, $\boldsymbol{\eta}$ being the associated parametrisation parameters. 

We can use the value function as a baseline for a better estimate of the gradient. Since $\left\langle V_{\pit}(\sigma) \, \nabla_{\boldsymbol{\theta}} \log \pit (\alpha\vert\sigma) \right\rangle = 0$, the gradient can be rewritten as 
\begin{equation}
\nabla_{\boldsymbol{\theta}} \bar{R}[\pi_{\boldsymbol{\theta}}] = \left\langle A_{\pit}(\sigma, \alpha)\, \nabla_{\boldsymbol{\theta}} \log \pit (\alpha\vert\sigma) \right\rangle,
\end{equation}
where $A_{\pit}(\sigma, \alpha) = Q_{\pit}(\sigma, \alpha) - V_{\pit}(\sigma)$ is the \textit{advantage function}. We furthermore use that the temporal-difference error of the value function is an unbiased estimate of the advantage function, namely
\begin{align}
	A_{\pit}(\sigma, \alpha) &= \langle \delta \rangle, \mbox{ with } \\
	 \delta &= R(\sigma_t,\alpha_t) - \bar{R}[\pit] + V_{\pit}(\sigma_{t+\Delta t}) - V_{\pit}(\sigma_t),\nonumber
\end{align}
leading to sample the performance-measure gradient as ${\nabla}_{\boldsymbol{\theta}} \bar{R}[\pi_{\boldsymbol{\theta}}] \approx \delta\,\nabla_{\boldsymbol{\theta}} \log \pit(\alpha_t, \vert, \sigma_t)$ and use this approximation to update the policy parameters. As for the gradient of the policy, we use the soft-max approximation~(\ref{eq:soft-max}) to write
\begin{align}
	\partial_{\theta_{ij}} \!\log\pit(\alpha\vert\sigma) &= \delta_{\Omega(\sigma),i} \delta_{\alpha, j} - \frac{1}{\mathcal{Z}_i} \mathrm{e}^{\theta_{ij}}\delta_{\Omega(\sigma),i} \nonumber\\ &= \delta_{\Omega(\sigma),i} \left[\delta_{\alpha, j} - \pit(j\vert \sigma)\right].
\end{align}
In practice we use an approximation of the value function $V_{\pit}(\sigma) \approx V_{\boldsymbol{\eta}}(\sigma) = \sum_i \eta_i \delta_{\Omega(\sigma),i}$ with parameters $\boldsymbol{\eta}\in\mathbb{R}^6$, in order to compute $\delta$. We trivially get $\partial_{\eta_i} V_{\boldsymbol{\eta}}(\sigma) = \delta_{\Omega(\sigma),i}$.

Summing up these expressions finally yields the procedure presented in Algorithm \ref{alg1}.
\begin{algorithm}[H]
  \caption{Policy gradient / Actor-Critic}
  \label{alg1}
  \begin{algorithmic}[1]
  \State Algorithm parameters: rates $\lambda_1, \lambda_2, \lambda_3$
  \State Initialize $\omega$, $\alpha$, $\bar{R}$ and the parameters $\boldsymbol{\theta}$ and $\boldsymbol{\eta}$
  \State Initialize the state $\sigma$ and the action $\alpha$
  \For{$n = 1, 2, \dots$}
      \State Take action $\alpha$ and evolve to the new state $\sigma'$
      \State Measure  reward $R$ and new observation $\omega' = \Omega(\sigma')$
      \State Select the next action $\alpha'\sim\pit(\cdot\,\vert\, \sigma')$
      \State Compute the TD error $\delta = R - \bar{R}+ \hat{V}_{\boldsymbol{\eta}}(\sigma') - \hat{V}_{\boldsymbol{\eta}}(\sigma)$
      \State $\bar{R} \gets \bar{R} + \lambda_1\Delta t\,\delta$
      \State $\boldsymbol{\eta}(\omega) \gets \boldsymbol{\eta}(\omega) + \lambda_2\Delta t\,\delta$
      \State $\boldsymbol{\theta}(\omega,\cdot) \gets \boldsymbol{\theta}(\omega,\cdot) +\lambda_3\Delta t\, \delta\, [\delta_{\alpha,\cdot} - \pit(\cdot\, \vert\,\sigma)]$
      \State $\omega \gets \omega'$, $\alpha \gets \alpha'$
  \EndFor
   \end{algorithmic}
  \end{algorithm}

We evaluate the performance of the actor-critic policy gradient algorithm in our navigation problem with 
the learning rates $\lambda_1 = 5\times10^{-7}\nu$ and $\lambda_2=\lambda_3=5\times10^{-5}\nu$. 
The results reported in Fig.~\ref{fig_actor_critic_displ}b show that the swimmer reaches a swimming velocity that is similar to that of the
$\varepsilon$-greedy $Q$-learning algorithm during forward motion periods.
However, unlike the $Q$-learning
algorithm which suffers from an enormous variability, the results obtained by actor-critic
are more consistent and stable, showing minimal variability across different realisations.
As seen in Fig.~\ref{fig_actor_critic_displ}a , the learning process of the swimmer shows the desired behaviour
as the swimming velocity systematically, albeit slowly, improves overtime. 
The evolution of the value function of the different observations (Fig.~\ref{fig_actor_critic_displ}c)
uncovers a sizeable amount of the story, it highlights how the 
swimmer initially learns to distinguish that tailwind is better than no significant wind,
which in turn is better than headwind. Much later in the process, it ends up learning the obvious (to us humans)
fact that being righty oriented is better than being wrongly oriented.
Eventually, as can be seen by the very end of this run, it starts to make more precise evaluations by starting
to learn that orientation is more important than the wind direction and magnitude.
This improvement, which is only reached at the end of the run, indicates that there is still significant potential
for further improvement in the swimmer's performance when using such an algorithm.

\begin{figure}[h]
\centering
\includegraphics[width=\columnwidth]{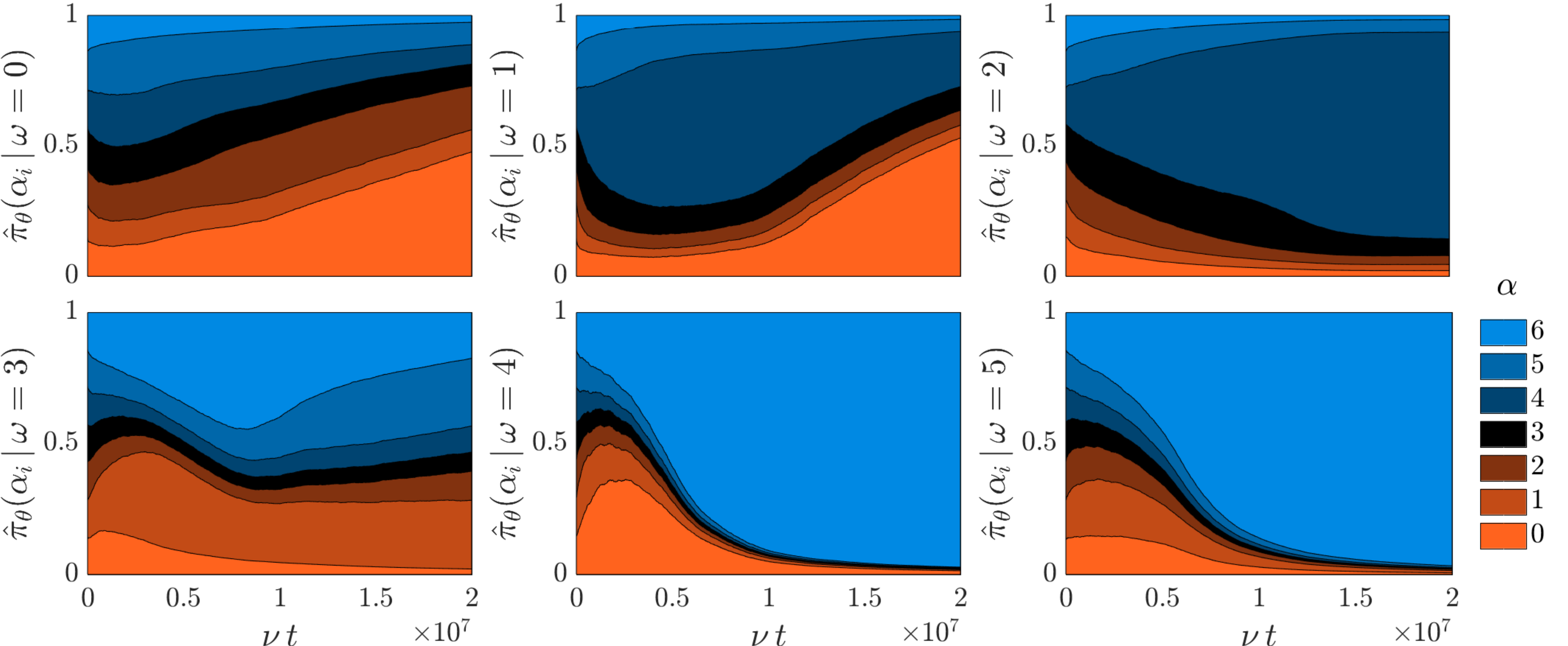}
\caption{\label{fig_actor_critic_policy} Time evolution of the approximated policies, shown for each value of the observation $\omega \in\{0..5\}$. The probability of choosing a given action $\alpha$ is shown as a coloured area.}
\end{figure}
Regarding the policy, as shown in Fig.~\ref{fig_actor_critic_policy}, the swimmer evolves and 
adapts its strategy over time in the course of the learning process. The policy starts from a random state where 
the swimmer is equally likely to choose any of the seven possible actions and thus basically being carried 
along by the flow. Over time, it learns to select with higher probabilities 
the actions that are more likely to lead to an 
improvement in its horizontal displacement. 
The swimmer, for instance, eventually discovers that action 6 is the most effective when it is 
oriented in the right direction and the wind is blowing in the right direction or not significant. 
This may seem obvious to us, but it took the swimmer a long time to figure it out. 
It is worth mentioning that this run for the actor-critic algorithm is longer by a factor of 10 compared 
to previous runs, but the performance of the swimmer still improves consistently, although at a slower
pace than in the early stages of the process.

All in all, the actor-critic algorithm presents a learning process that is more stable and 
consistent across runs compared to $Q$-learning. This stability leads to a policy that is 
incrementally improved during learning, resulting in the desired feature of improved performance 
over time. However, despite its consistent learning process, the swimmer's performance 
achieved through the actor-critic algorithm falls short of the results obtained with the naive strategy and 
requires substantial computational resources if it is to surpass it.

\subsection{Competitive $Q$-learning}
\label{subsec:competitive}
We have seen in previous subsections that various methods of reinforcement learning fail to provide a 
satisfactory answer to our optimisation problem.  On the one hand, the idea of bluntly applying methods 
designed for Markovian systems, such as $Q$-learning, suffers non-convergence. On the other hand, 
approximation approaches, which were in principle developed to tackle partial observability, face 
issues related to an extremely slow convergence, if any, making their use ineffective or even 
prohibitive. Moreover, all the policies that emerged as intermediate or asymptotic outcomes of these 
trials, were found to be significantly less performant than the naive strategy introduced in 
Sec.~\ref{subsec:naive}. We interpret these difficulties as a consequence of the rather brusque 
manner with which we have projected the high-dimensional set of swimmer's configurations onto a very 
small number of discrete observables.  Such a reduction of information triggers the chaoticity of the 
system, including of the learning process and this explains the sensitivity of our results to both 
the method and the particular chosen trajectory during iterative optimisation procedures.

In light of these observations, we present here a new perspective. In place of searching for a single efficient policy that would possibly outperform the naive strategy, we propose to construct a set of admissible policies. To make such a construction as easy as possible, we consider 200 different realisations of deterministic $Q$-learning 
(with a vanishing exploration parameter i.e., $\varepsilon=0$) that are obtained by choosing randomly the 
initial orientation of the swimmer. Each realisation of the learning algorithm is run in parallel to 
the others for a time $t=2\times10^5\nu^{-1}$. After this duration, the deterministic $Q$-learning 
algorithm has in all cases stabilised to a given policy, even if the entries of the $Q$-table have 
not converged.  This evolution to a fixed policy is a result of our decision to eliminate exploration by setting $\varepsilon = 0$.  The 200 policies obtained 
this way are then used to construct our admissible set.

\begin{figure}[t]
\includegraphics[width=.55\columnwidth]{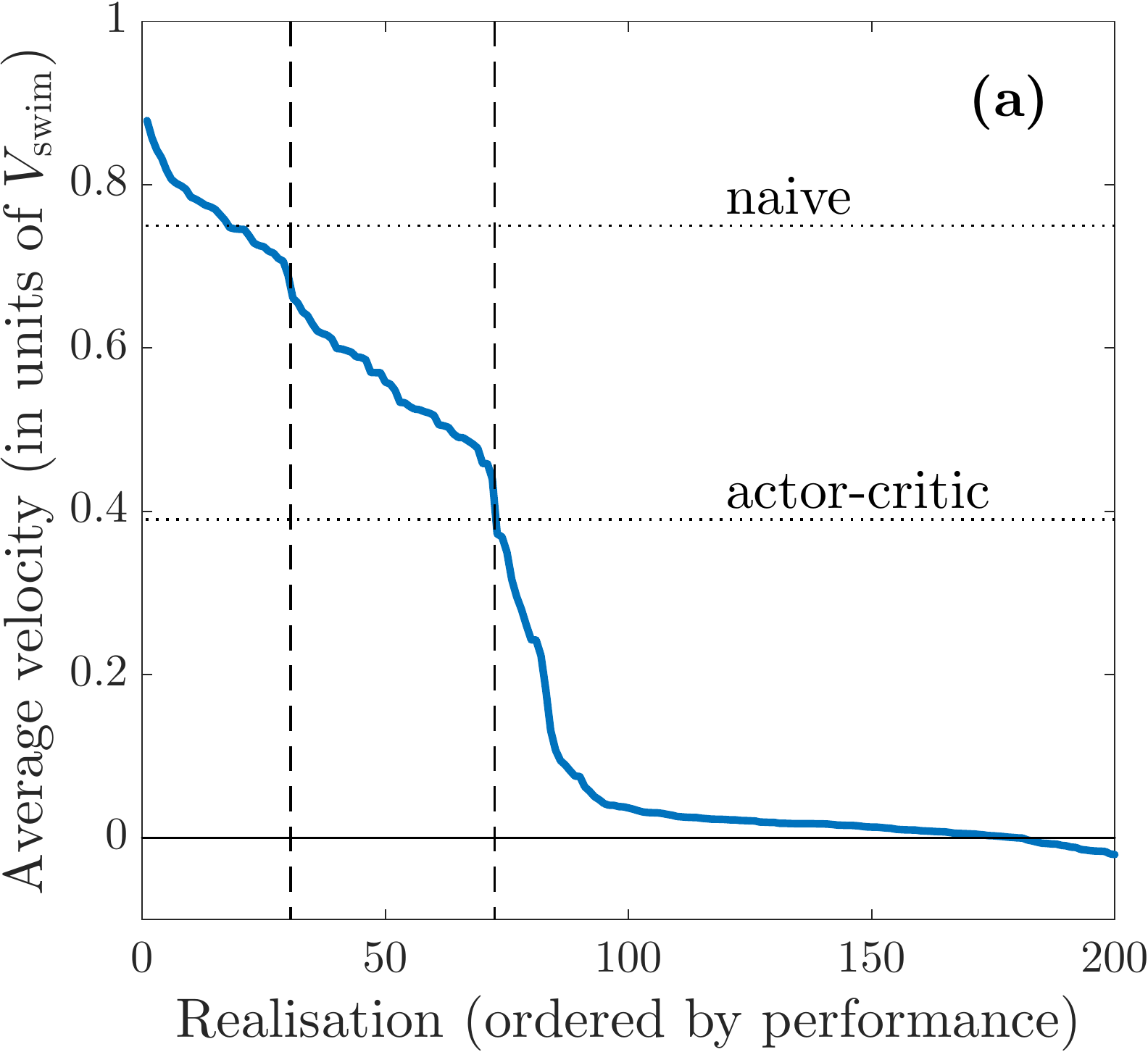}\\[8pt]
\includegraphics[width=.7\columnwidth]{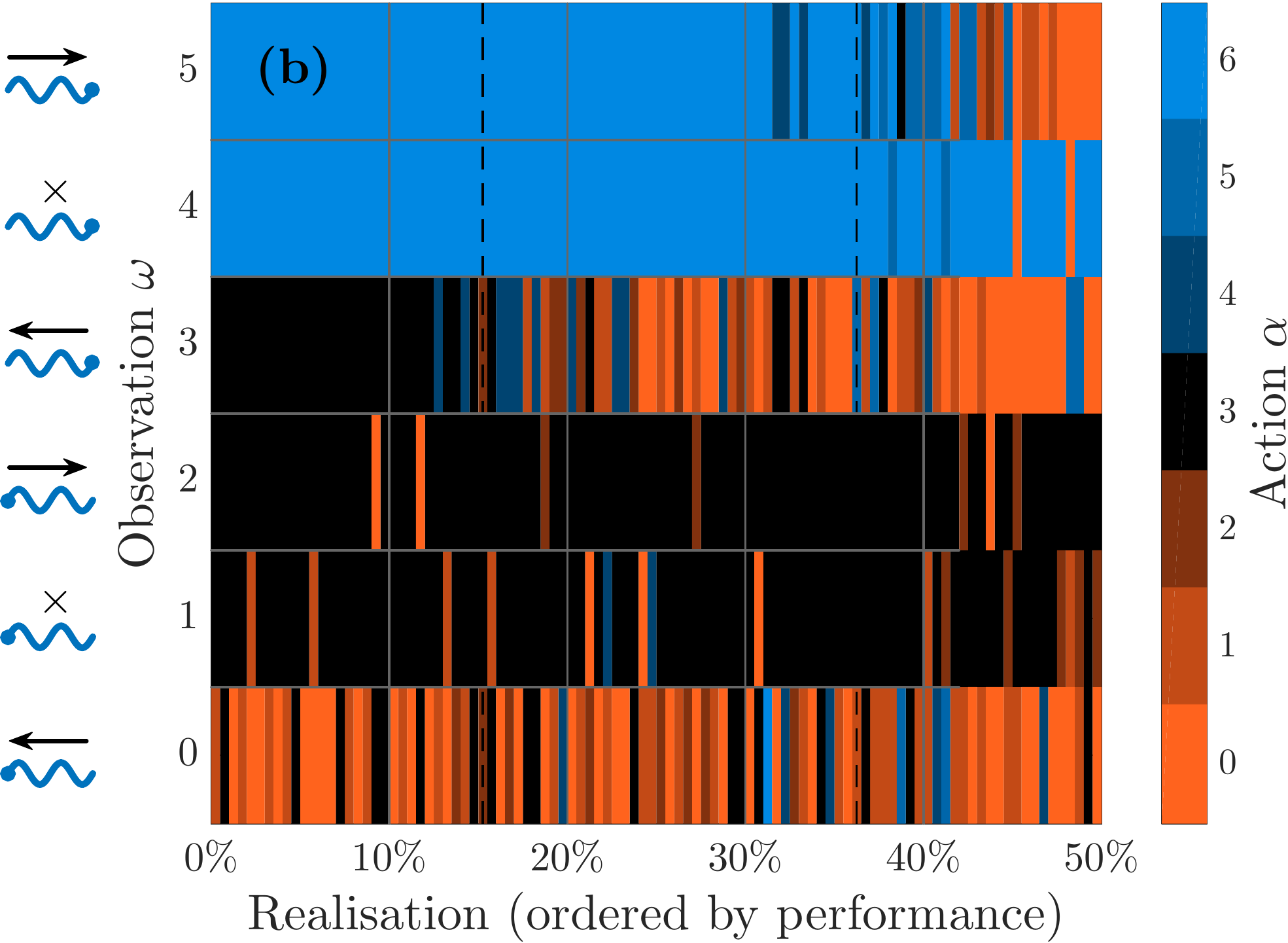}
\caption{\label{fig_simpleQ} Results of $200$ realisations of deterministic $Q$-learning. \textbf{(a)}~Average swimming speeds, ordered by decreasing performance. The two dotted lines show the velocities of the naive policy and that obtained with the Actor-Critic algorithm; the two dashed vertical lines mark quasi discontinuities in the swimmers performance. \textbf{(b)}~Strategies that significantly lead to the swimmer's displacement, again ordered from the most performant to the least. The two dashed vertical lines mark the same change of behaviour as on the left panel.}
\end{figure}
Figure~\ref{fig_simpleQ}a shows the asymptotic velocity $\delta_\tau \bar{x}_1 / \tau$ attained by these 200 instances of $Q$-learning, ordered by decreasing efficiency.  One observes roughly three classes of policies, which are separated in the plot by vertical dashed lines. The top 15\% perform better, or comparably to the naive strategy. The next 15 to 37\% realisations overcome the strategy obtained by the actor-critic algorithm and give a reasonable displacement of the swimmer. Finally, the worse 63\% do not yield any significant displacement. These three regimes are separated by abrupt jumps in the average velocity. As evidenced from Fig.~\ref{fig_simpleQ}b, they correspond to significant changes in the corresponding policies. The top 15\% policies clearly belong to the same category as the naive strategy. They are all prescribing a vigorous vertical undulation ($\alpha=6$) when the swimmer is favourably oriented and feels no headwind ($\omega=4$ and $5$). They  essentially recommend to stop swimming ($\alpha = 3$) for a right orientation and a headwind ($\omega = 3$) or when the swimmer is directed the wrong way and experiences a headwind ($\omega = 1$ and $2$). They favour horizontal undulations ($\alpha=0$ and $1$) or to stop swimming ($\alpha = 3$) when the swimmer is wrongly oriented with the flow blowing to the left. These top strategies mostly differ by the actions chosen for $\omega=0$, 1, and 2. The separation with the second family of policies is clear from Fig.~\ref{fig_simpleQ}b: It corresponds to a change in the action performed for $\omega=3$, from stopping swimming to undulating in the horizontal direction. The second change separating the second and third categories is as clear: The policies stop there prescribing a vertical undulation in the favourable configuration $\omega=6$.

When looking with more detail at the 15\% top-ranked outcomes of $Q$-learning, one notices 
that the corresponding policies are rather few. They indeed form a set of five admissible 
policies whose performances are very similar and overtake each other depending on the realisation 
of the algorithm. In addition to the naive strategy, which can be written as 
$\alpha_\pi = [3,3,3,3,6,6]$ where the $i$-th element of the array corresponds to the action 
$\alpha_\pi(\omega)$ followed when $\omega=i$, the other four policies are 
$\alpha_\pi = [0,3,3,3,6,6]$, $[1,3,3,3,6,6]$, $[3,3,0,3,6,6]$, and $[0,1,3,3,6,6]$.  
Notice that none of these five policies emerged, neither in an intermediate stage, nor asymptotically, 
in the various trials of reinforcement learning in previous subsections. We select these five 
strategies to define a set of admissible deterministic policies that are potential solutions to our 
optimal navigation problem. In the next section, we address with more detail their actual performance and 
robustness when varying the physical settings in our system.

\section{Performance and robustness of the admissible policies}
\label{sec:robustness}

\subsection{Long-term statistics}
\label{subsec:stat}
\begin{figure}[b]
\includegraphics[width=.65\columnwidth]{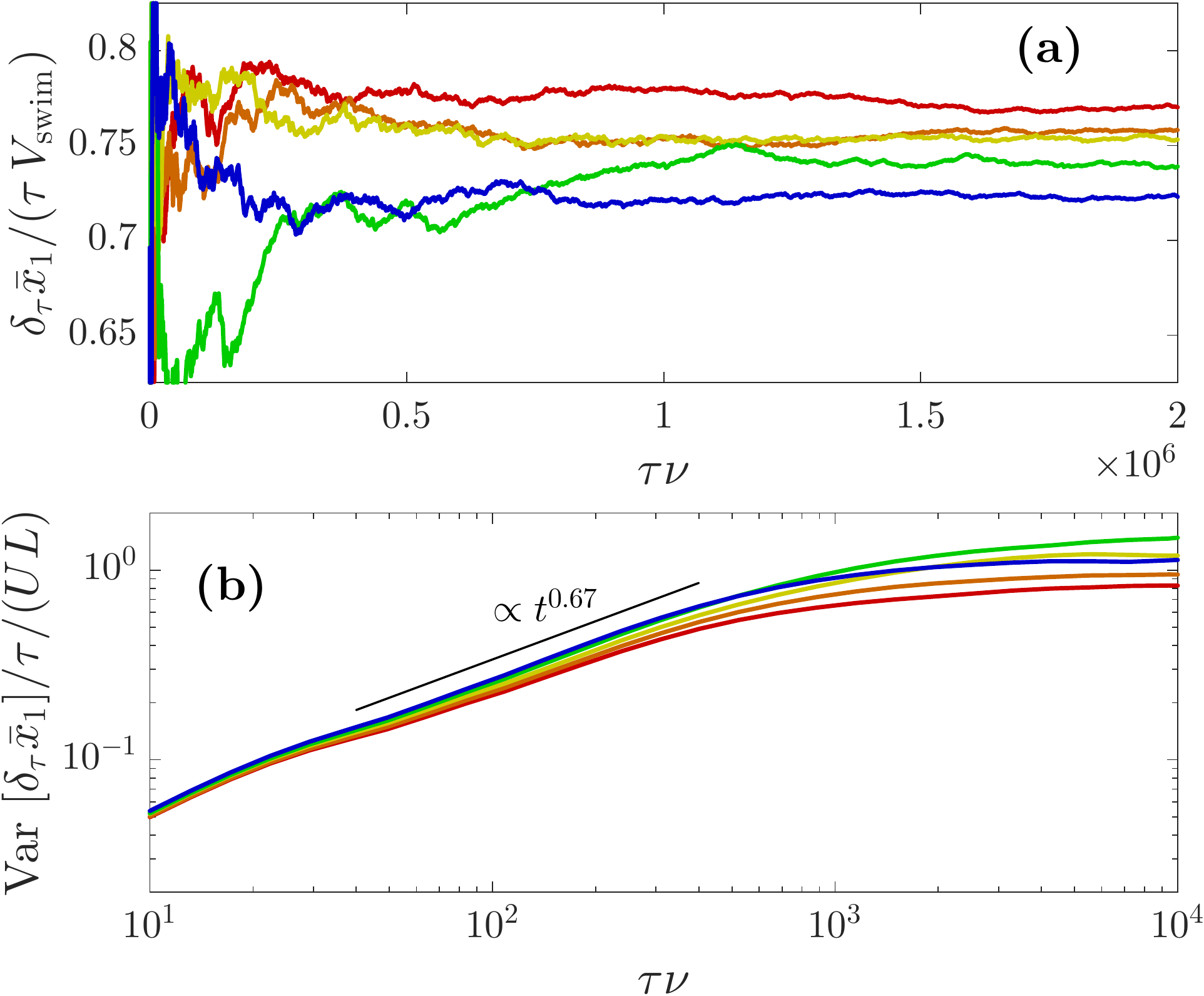}\\[8pt]
\includegraphics[width=.55\columnwidth]{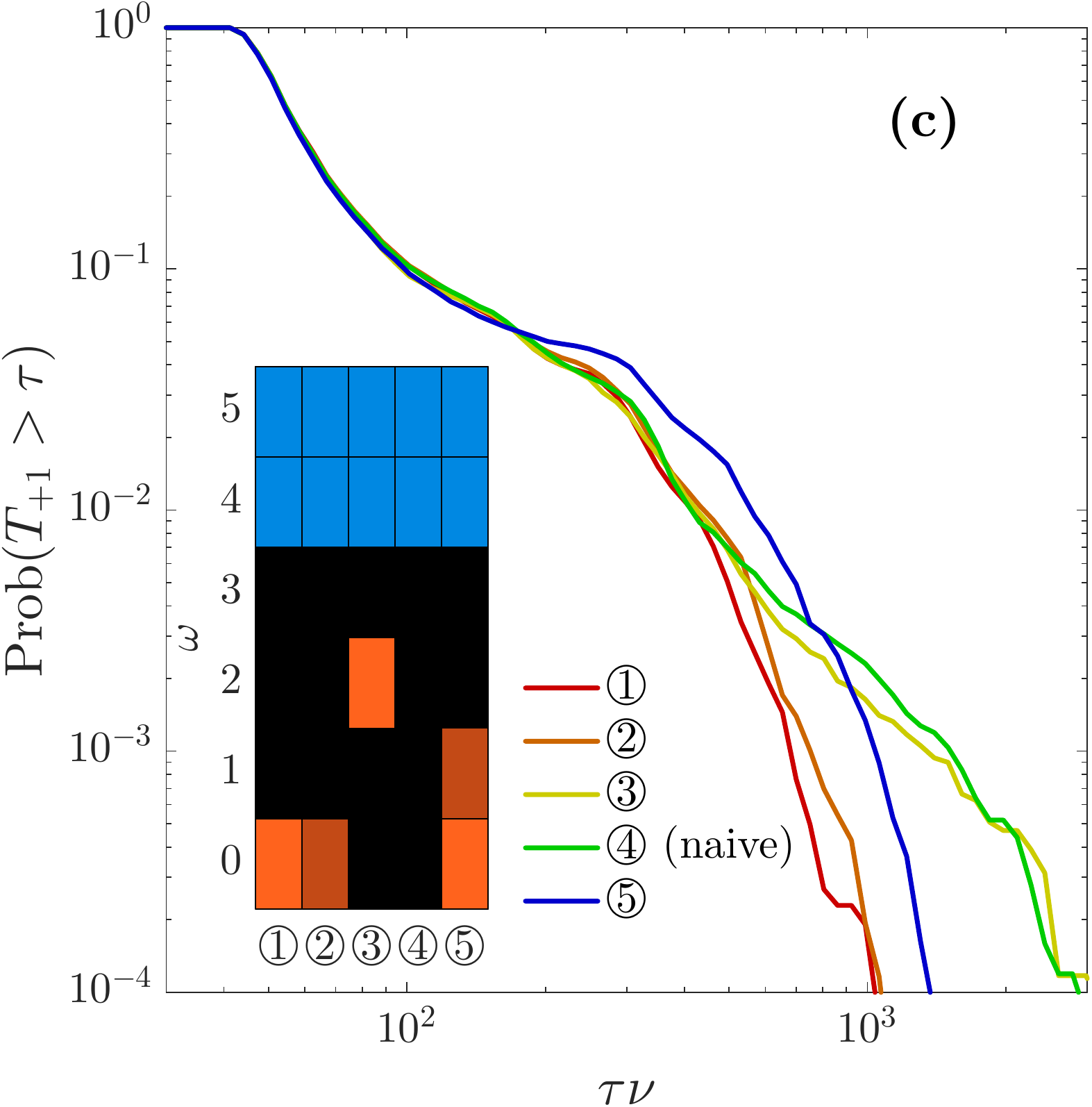}
\caption{\label{fig_compare_strateg} Comparison of the admissible strategies \circl{1} to \circl{5} ---\,as depicted in the inset of panel \textbf{(c)}\,--- again evaluated for $\mathcal{F}=15$, $U=0.025\,\ell\nu$, $\ell/L = 1$, $u_0/U = 0.2$, and $A_0 = 0.08$.  \textbf{(a)}~Time-averaged horizontal velocity $\delta_\tau \bar{x}_1/\tau$ as a function of the time lag $\tau$. \textbf{(b)}~Effective diffusion coefficient obtained from the variance of $\delta_\tau \bar{x}_1$.  \textbf{(c)}~Complementary cumulative distribution function of the first passage time $T_{+1}$ from $x_1=j\,L$ to $x_1 = (j+1)L$.}
\end{figure}

We here provide details on the performance of the five admissible strategies obtained from competitive realisations of deterministic $Q$-learning in Sec.~\ref{subsec:competitive}. Figure~\ref{fig_compare_strateg}a shows the time evolution of the velocity $\delta_\tau \bar{x}_1/\tau$ along trajectories that each follow one of the selected policies (velocities are there expressed in units of the swimming speed $V_{\rm swim}$ in absence of flow). Unambiguous differences in the performance of the different policies are only visible for $\tau\gtrsim 10^6\nu^{-1}$, the shorter time lags being essentially dominated by violent fluctuations of the displacement. This very slow convergence of the time averages along the swimmer dynamics can clearly be an important source of difficulties when using iterative optimisation algorithms. We hereafter label these trajectories from \circl{1} to \circl{5} using their efficiency ranking. The naive policy is \circl{4} and the diagrams of the other admissible policies are shown in inset of Fig.~\ref{fig_compare_strateg}c.

The variances of the displacement over a time $\tau$ evaluated for the five policies are shown in Fig.~\ref{fig_compare_strateg}b. We have here divided it by the time lag $\tau$ in order to measure effective coefficients of diffusion. One observes almost the same ordering of trajectories (except for  \circl{5}), suggesting that good performance goes together with weaker fluctuations. All curves saturate to a plateau at large times, indicating a long-term diffusive regime of the horizontal displacement about its average, as already observed for the naive strategy in Sec.~\ref{subsec:naive}. The asymptotic value gives an estimate of the associated coefficient of diffusion. For all policies, it is of the order of the fluid-flow units, namely $UL$, which is itself of the order of the displacement units $\simeq V_{\rm swim}\ell$. This means that, on a time $L/V_{\rm swim}$ needed by the swimmer to travel across a cell, it typically diffuses over a distance equal to the cell size $L$ itself. This strong diffusion accounts for the observed slow convergence of the average velocity. The order of magnitude of diffusion exactly corresponds to a finite contribution from trapping. It indicates that on a time $L/V_{\rm swim}$, the swimmers can remain with a finite probability in the same cell rather than moving to the next one.

These considerations become much clearer when measuring the probability distribution of the time $T_{+1}$ that the swimmer needs to travel from one cell to the next adjacent one. The complementary cumulative distribution functions obtained for the five policies are shown in Fig.~\ref{fig_compare_strateg}c. All curves almost collapse on the top of each other, up to times corresponding to hundreds of undulatory beats. Admissible policies therefore differ little in their ability to move the swimmer when its conditions are standard.  Nonetheless, marked difference are found in the tails of the distributions, which are sampling trapping events. The two most performant policies (\circl{1} and \circl{2}) are associated to lesser probabilities of getting a long transition time $T_{+1}$. This can be interpreted as a consequence of the horizontal undulation that both policies recommend when the swimmer is wrongly oriented with a negative fluid velocity ($\omega=0$). Such a choice apparently makes a difference with the two next policies (\circl{3} and \circl{4}) that both display a fatter tail in the distribution of $T_{+1}$. For these two policies, the swimmer stops undulating when in such a configuration. Finally, policy~\circl{5}, which is beaten by the four others, shows a higher probability at smaller values of $T_{+1}$, possibly indicating that it is more likely to bring about trapping, even if  swimmers can then escape faster.

\subsection{Robustness with respect to the physical parameters}
Here we address the performance of admissible policies by varying the physical properties of the swimmer. We have performed a set of numerical simulations where we alternatively vary the size ratio $\ell/L$, the swimmer flexibility $\mathcal{F} =  (\zeta\nu/K)^{1/4}\ell$, or the velocity ratio $U/V_{\rm swim}$, while keeping constant the two other parameters. We estimated from these simulations average swimming speeds by monitoring again the asymptotic displacements of the swimmers.

\begin{figure}[t]
\includegraphics[width=\columnwidth]{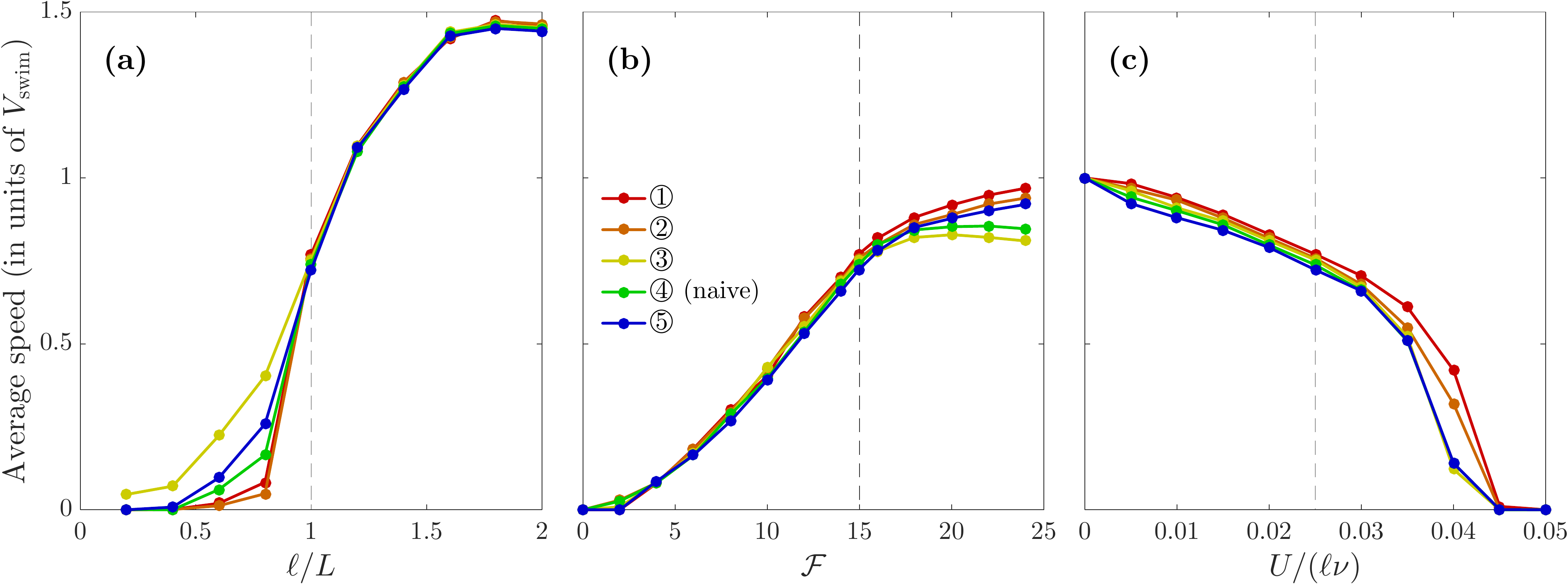}
\caption{\label{fig_phys_param} Robustness of the admissible strategies \circl{1} to \circl{5} when varying the swimmers physical parameters. All results where obtained for $u_0/U = 0.2$ and $A_0 = 0.08$. \textbf{(a)}~Average swimming speed as a function of the ratio between the swimmer length $\ell$ and the flow length scale $L$ (for $\mathcal{F}=15$ and $U=0.025\,\ell\nu$ both fixed). \textbf{(b)}~Same, as a function of the swimmer flexibility $\mathcal{F} =  (\zeta\nu/K)^{1/4}\ell$ and the flow length scale $L$ (for $\ell/L=1$ and $U=0.025\,\ell\nu$). \textbf{(c)}~Same as before,  varying this time the fluid flow velocity $U$ (for $\ell/L=1$ and $\mathcal{F}=15$). On each panel, the vertical dashed line shows the parameter value used in earlier sections.}
\end{figure}

Figure~\ref{fig_phys_param}a shows the performance of five policies obtained at varying the length $\ell$ of the swimmer. We find that dependence upon policy is only visible for swimmers that are sufficiently small compared to the cell size, whereas at larger sizes, the five policies perform comparably well.  One indeed observes for $\ell\lesssim 0.8\,L$ that the performance ranking of policies is completely shuffled. The swimmers following the otherwise efficient policies \circl{1} and \circl{2} barely move towards $x_1>0$, while the best performance is achieved by \circl{3}. This can be understood by the fact that tumbling and trapping completely change their natures for short swimmers (or equivalently large-scale fluid inhomogeneities).  The action of trying to escape by a vigorous vertical swim seems hence less efficient than just stopping to swim and waiting for being conveyed by the flow in a more favourable region. At larger swimmer sizes ($\ell\gtrsim 0.8\,L$), the ranking between the various policies seems almost independent of $\ell/L$, even if the various policies seem to asymptotically perform similarly. The swimming speed seems to saturate for $\ell\gtrsim 1.8\,L$. This due to the fact that long swimmers are very unlikely to get tumbled by the flow, so that what only matters are the actions performed for observations $\omega=3$, $4$, and $5$ and they are identical for the five admissible policies.

Figure~\ref{fig_phys_param}b shows dependence upon flexibility. The various policies perform equally well for  rigid swimmers (small $\mathcal{F}$). In that case, they are almost never bent, nor buckled by the flow. This prevents trapping, and thus does not allow the various policies to display any differences in performance. At the same time and as seen in Sec.~\ref{subsec:undulatory}, much energy is dissipated by the elastic forces, hindering efficient swimming motions. The differences  between the various strategies is however much more visible for flexible swimmers (large $\mathcal{F}$).  Policies that efficiently prevent long-term traps (\circl{1}, \circl{2} and \circl{5}) stand clearly out from the two others. This divergence is promoted by flexibility, because the swimmers are more and more likely to get trapped when $\mathcal{F}$ increases.

Finally, figure~\ref{fig_phys_param}c show results obtained when varying the amplitude $U$ of the outer fluid flow. For all policies, the average horizontal velocity decreases from the swimming speed in the absence of flow ($U=0$) to very small values for strong fluid flows. None of the admissible policies lead to any significant displacement of the swimmers for fluid velocities exceeding $\simeq 0.045\,\ell\nu \simeq 2.5\,V_{\rm swim}$. It seems from our measurements that the performance ranking between the five policies does not depend on $U$.

\subsection{Tests in two-dimensional unsteady flows}
\begin{figure}[h]
\centering\begin{minipage}{0.35\columnwidth}
\includegraphics[width=\textwidth]{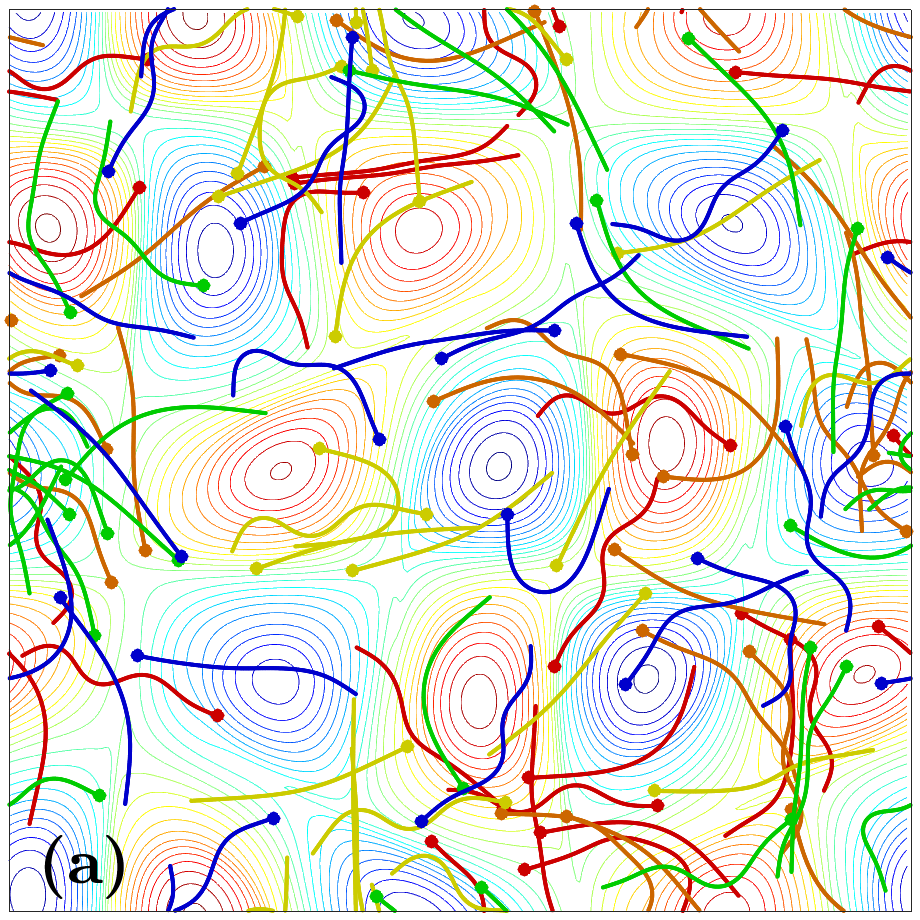}\\[4pt]
\includegraphics[width=\textwidth]{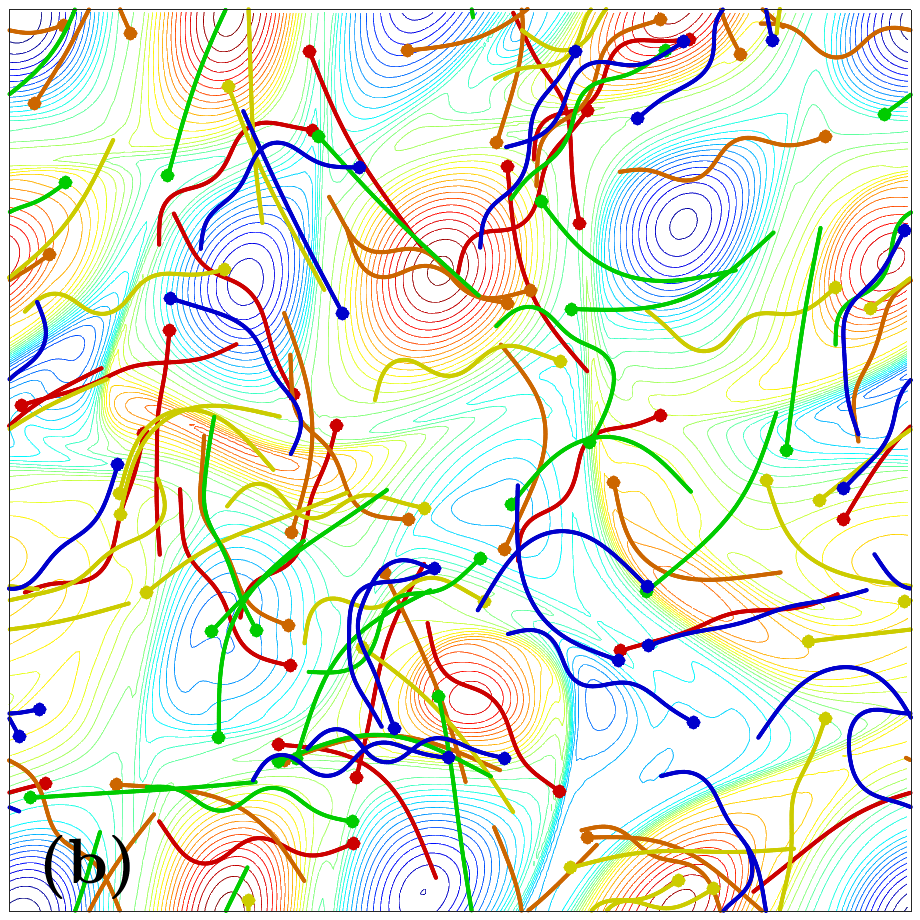}
\end{minipage}
\hfill
\begin{minipage}{0.6\columnwidth}
\includegraphics[width=\textwidth]{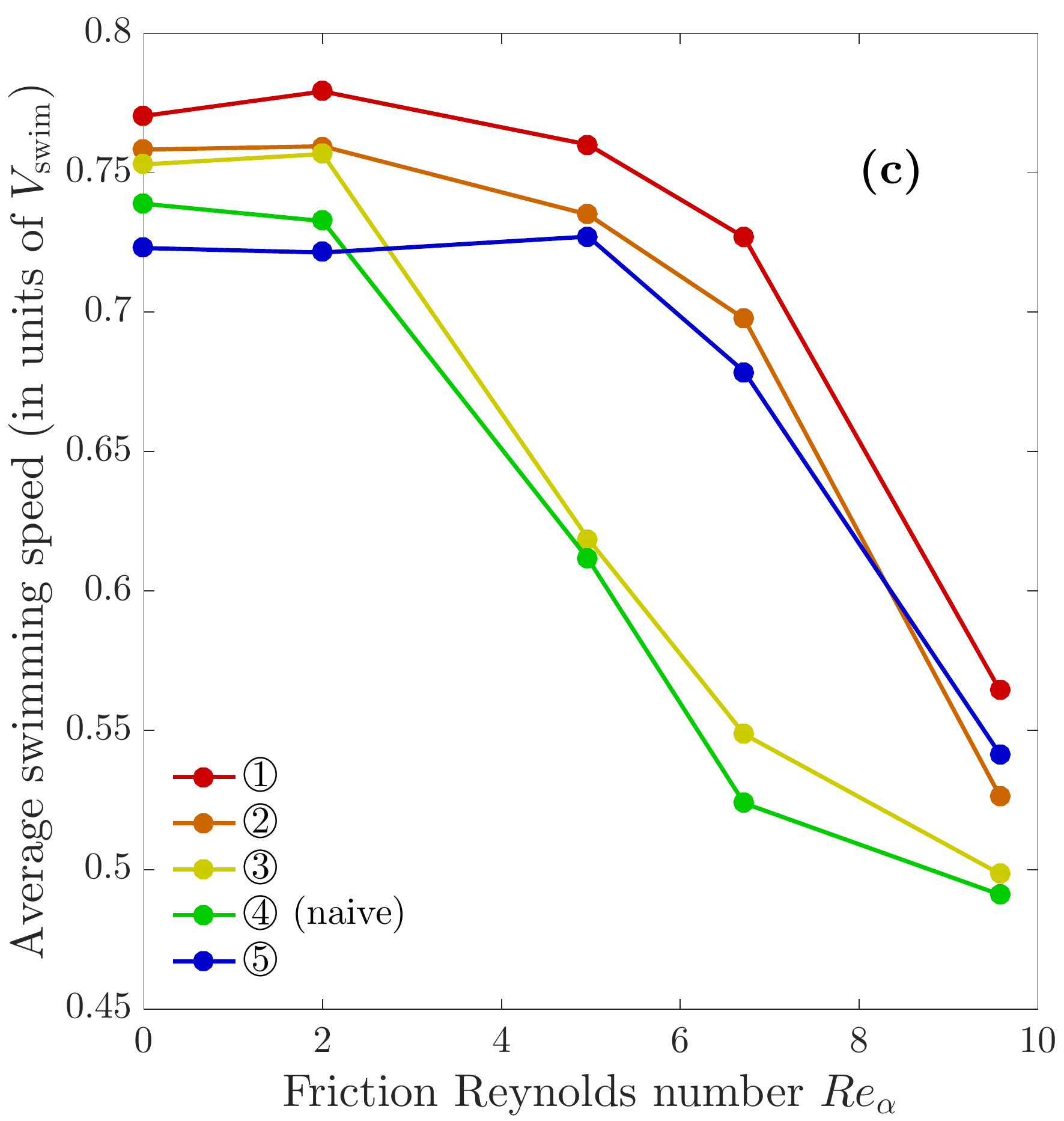}
\end{minipage}
\caption{\label{fig_turb_flow} Swimmers immersed in a non-steady flow that follow the five admissible policies. Left panels: snapshot of the fluid vorticity $\omega = \partial_1u_1-\partial_2u_1$ (contour lines in background), together with the instantaneous position of swimmers coloured according to the policy they follow, for \textbf{(a)} $Re_\alpha\simeq 2$ and \textbf{(b)} $Re_\alpha\simeq9.5$.  Right panel \textbf{(c)}~Average swimming speed as a function of the friction Reynolds number $Re_\alpha$ for the five admissible policies, as labeled.}
\end{figure}

To assess further the robustness of the proposed policies, we consider now the case where the swimmers are moving in a more realistic stationary flow that solves the incompressible Navier--Stokes equations. The fluid velocity field, in place of being a  steady cellular flow, is now a solution of
\begin{align}
&\rho_{\rm f}\left[\partial_t \bm u + \bm u\cdot\nabla\bm u\right] = -\nabla p + \mu\nabla^2\bm u -\alpha\bm u + \nabla^\perp F, \nonumber\\ &\nabla\cdot\bm u = 0,
\end{align}
where $\rho_{\rm f}$ is the fluid mass density, assumed constant, $\mu$ is its dynamic viscosity, $\alpha$ is a friction coefficient accounting for the two-dimensional confinement of the flow, and $\nabla^\perp F$ is an external incompressible force that maintains the flow in motion.  We choose the stationary cellular force $F = (\alpha U L/\pi)\cos(\pi x_1/L)\,\cos(\pi x_2/L)$, with a forcing amplitude $U$ and a spatial period $L$ that set the large velocity and length scales of the flow. The dynamics then depends upon two non-dimensional parameters, the viscous Reynolds number $Re_\mu = \rho_{\rm f}U\,L/\mu$, which balances inertia and viscous dissipation, and the friction Reynolds number $Re_\alpha = \rho_{\rm f}U/(L\alpha)$, which balances inertia and friction.  Depending on them, the flow might bifurcate between different regimes~\cite{perlekar2010turbulence,michel2016bifurcations}. We assume $Re_\mu\gg 1$, so that viscous dissipation acts only at small scales, possibly yielding a direct turbulent cascade of enstrophy.  By contrast $Re_\alpha$ is used as a control parameter. With this choice one recovers when $Re_\alpha \ll 1$ the stationary cellular flow that has been previously considered. When $Re_\alpha$ increases, the flow transitions to a turbulent regime where it is unsteady and chaotic. Illustrations of the associated vorticity fields are given in Fig.~\ref{fig_turb_flow}a and b. 

We have performed several numerical simulations of the fluid velocity at varying the friction Reynolds number $Re_\alpha$. We used a pseudo-spectral solver with $256^2$ collocation points, second-order Runge--Kutta time marching, and implicit integration of the linear viscous and friction terms. The velocity is assumed $2\pi$-periodic and we set $L = \pi/2$. Various swimmers are embedded in the flow (20 for each policy from \circl{1} to \circl{5}) and we monitor their progression toward $x_1>0$.  In all simulations, both physical and navigation parameters are kept the same as in Sec.~\ref{subsec:stat}, namely $\mathcal{F}=15$, $U=0.025\,\ell\nu$, $\ell/L = 1$, $u_0/U = 0.2$, and $A_0 = 0.08$.  The average horizontal speed of such swimmers is reported in Fig.~\ref{fig_turb_flow}b as a function of the friction Reynolds number. At low $Re_\alpha$, one recovers the same performance ranking as previously observed in stationary cellular flows. However, for $Re_\alpha>4$, the flow develops a chaotic regime characterised by open streamlines with rather strong jets where the swimmer might be entrained in an inappropriate direction. The performance of policies \circl{3} and \circl{4} drops down significantly, while the other policies continue to operate relatively well. This dissimilarity can be explained by the contrasting responses to trapping observed in Sec.~\ref{subsec:stat}. Policies \circl{1}, \circl{2} and \circl{5} have in common to promote a horizontal undulation when the swimmer is wrongly oriented with a headwind. This action allows the swimmer to move transversally and escape strong persistent jets that otherwise sweep it toward $x_1<0$. This makes apparently a noticeable difference at intermediate values of $Re_\alpha$.

\section{Concluding remarks}
\label{sec:conclusion}

We have studied in this paper the question of optimising the displacement of undulatory, deformable micro-swimmers evolving in a prescribed, non-homogeneous outer flow. Our physical model imposes close links between the macroscopic displacement of the swimmer and its microscopic active deformation that induces motility.  This clearly differs from standard optimal-navigation problems, which generally assume a scale separation between these two mechanisms that is sufficiently large to consider them independent from each other.  We used reinforcement-learning methods to address this problem, trying constantly to interpret the outcomes of our approaches in terms of the underlying physics. An important message that we want to convey is the necessity of determining the relevant physical timescales of the problem. This leads not only to choosing appropriate hyper-parameters of the learning schemes, but also to estimating and understanding their convergence rate.

In our settings, the swimmer's configurations form a high-dimensional set from which we arbitrarily decided to exploit only very partial information. However, these settings happened to constitute a clear instance where the prescription of only a  limited knowledge of the state of the agent has drastic impacts on the optimisation of the decision process.  We have tested several methods, ranging from simple $Q$-learning to more sophisticated approximation methods. All these trials lead to prohibitively-long convergence times, if not infinite.  To our opinion, this is due to the fact that the information on the swimmer's configuration is so coarse that our problem deviates in a significant manner from the usual Markovian framework. This, combined with chaotic dynamics, leads to tremendous fluctuations with respect to initial data that jeopardise the outcomes of reinforcement-learning procedures.  The combination of a very-partially observable character with a highly-chaotic nature of the system is certainly a feature shared by many other practical decision problems. It would be of significant interest to formalise better such connections by evaluating for instance the stability and ergodicity of the global dynamical system defined as the union of the iterative learning procedure and the underlying dynamics.

Despite these difficulties, we have proposed an alternative approach based on concurrent realisations of reinforcement learning. Instead of limiting the optimisation procedure to searching for a unique satisfactory approximation of the optimal policy, we shifted our objective to constructing an almost-comprehensive set of admissible strategies whose performance and robustness might be assessed subsequently.  The case we have considered is particularly rich, while remaining tractable. The set of admissible strategies was obtained in a quite simple manner by running different instances of deterministic $Q$-learning, whose results proved to be particularly sensitive to the specific initialisation of the algorithm. Moreover, the set constructed this way reduces to only five different admissible policies, making rather easy any systematic assessment on their efficiencies. Still, as demonstrated in Sec.~\ref{sec:robustness}, the performance of each of these policies can appreciably vary when changing the physical parameters of the swimmer or the type of fluid flow in which it is immersed. Such a systematic investigation would have been impossible if one had to solve, for each setting, an expensive optimisation problem.

Finally, let us stress that most of the difficulties we faced could be stemming from the arbitrary choice of limited observables and actions that we considered in the decision process. The motivation for such a prescription was mainly coming from practical applications. In general, the amount of accessible information and of possible manoeuvres is strongly constrained by the design, cost, and efficiency of sensors and engines that equip a micro-robot, or by the primitive nature of the micro-organisms in consideration. However, it could largely be that the observables and actions that we have chosen are not enough for this physical model and the associated navigation problem. It would thus be interesting to repeat this analysis and the reinforcement-learning trials by adding, at both ends of the decision process, encoder-decoder neural networks that would automatically extract and redistribute the relevant information. Interpreting the encoded information could be highly pertinent to the design of optimal sensors and actuators and their implementation in practical applications. 

\

\section*{Acknowledgments}
The authors are grateful to the OPAL infrastructure from Universit\'{e} C\^ote d’Azur for providing computational resources. This work received support from the UCA-JEDI Future Investments funded by the French government (grant no.\ ANR-15-IDEX-01) and from the Agence Nationale de la Recherche (grant no.\ ANR-21-CE30-0040-01).

\bibliography{biblio}

\end{document}